\documentclass[twocolumn,trackchanges]{aastex631}

\usepackage{verbatim}
\usepackage{amsmath}
\usepackage{savesym}
\savesymbol{tablenum}
\usepackage{siunitx}
\restoresymbol{SIUNITX}{tablenum}

\submitjournal{ApJ}

\shorttitle{New mass estimates for massive binary systems}
\shortauthors{Fullard et al.}
\graphicspath{{./}{figures/}{figs/}{SLIPfigs/}}

\begin{document}

\title[]{New mass estimates for massive binary systems: a probabilistic approach using polarimetric radiative transfer}

\author[0000-0001-7343-1678]{Andrew G. Fullard}
\correspondingauthor{Andrew G. Fullard}
\email{E-mail: fullarda@msu.edu}
\affiliation{Department of Physics and Astronomy, Michigan State University, East Lansing, Michigan 48824 USA}
\author[0000-0003-3615-9593]{John T. O'Brien}
\affiliation{Department of Physics and Astronomy, Michigan State University, East Lansing, Michigan 48824 USA}
\author[0000-0002-0479-7235]{Wolfgang E. Kerzendorf}
\affiliation{Department of Physics and Astronomy, Michigan State University, East Lansing, Michigan 48824 USA}
\author[0000-0002-4022-1874]{Manisha Shrestha}
\affiliation{Astrophysics Research Institute, Liverpool John Moores University, Liverpool Science Park IC2, 146 Brownlow Hill}
\author[0000-0003-1495-2275]{Jennifer L. Hoffman}
\affiliation{Department of Physics and Astronomy, University of Denver, Denver, CO 80210 USA}
\author[0000-0002-7204-5502]{Richard Ignace}
\affiliation{Department of Physics and Astronomy, East Tennessee State University, Johnson City, Tennessee 37614 USA}
\author[0000-0003-4418-4916]{Patrick van der Smagt}
\affiliation{Machine Learning Research Lab, Volkswagen Group, Munich, Germany}
\affiliation{Faculty of Informatics, E\"otv\"os Lor\'and University, Budapest, Hungary}



\begin{abstract}
Understanding the evolution of massive binary stars requires accurate estimates of their masses. This understanding is critically important because massive star evolution can potentially lead to gravitational wave sources such as binary black holes or neutron stars. For Wolf-Rayet stars with optically thick stellar winds, their masses can only be determined with accurate inclination angle estimates from binary systems which have spectroscopic $M \sin i$ measurements. Orbitally-phased polarization signals can encode the inclination angle of binary systems, where the Wolf-Rayet winds act as scattering regions. 

We investigated four Wolf-Rayet + O star binary systems, WR 42, WR 79, WR 127, and WR 153, with publicly available phased polarization data to estimate their masses. To avoid the biases present in analytic models of polarization while retaining computational expediency, we used a Monte Carlo radiative transfer model accurately emulated by a neural network. We used the emulated model to investigate the posterior distribution of parameters of our four systems. Our mass estimates calculated from the estimated inclination angles put strong constraints on existing mass estimates for three of the systems, and disagrees with the existing mass estimates for WR 153. We recommend a concerted effort to obtain polarization observations that can be used to estimate the masses of Wolf-Rayet binary systems and increase our understanding of their evolutionary paths.
\end{abstract}






\section{Introduction}
Massive stars enrich the interstellar medium with heavy elements via stellar winds and supernova explosions. Their high-mass remnants, such as neutron stars and black holes, have been detected with gravitational wave observations \citep[e.g.][]{abbott_astrophysical_2016, abbott_multi-messenger_2017}. The evolutionary paths that massive stars take to reach different types of supernovae, or even core collapse without a luminous event, are strongly dependent on their masses and mass-loss rates \citep{puls_mass_2008, langer_presupernova_2012, abbott_astrophysical_2016, woosley_evolution_2019}.

A large fraction of sufficiently massive (18--80 M$_\odot$) individual stars are expected to evolve to the Wolf-Rayet (WR) stage \citep{sander_galactic_2019}, which is characterised by strong emission line spectra and high mass-loss rates \citep[$\sim$$10^{-5}$\,M$_\odot\,$yr$^{-1}$;][]{crowther_physical_2007}. However, the optically thick winds of WR stars make it impossible to measure their masses using $\log g$. Fortunately, massive stars typically occur in binary systems \citep{mason_high_2009, sana_southern_2014} so it is possible to obtain mass estimates using orbital models. Massive binaries can evolve into compact binaries that emit gravitational waves \citep{abbott_gw170817_2017}. Aside from using binaries for mass estimation, binary interactions can have a strong effect on the pre-supernova masses and structure of massive stars \citep[e.g.][]{laplace_different_2021} and more than 50\% of massive binaries interact during their lifetime \citep{sana_binary_2012,sana_vlt-flames_2013}. Mass transfer as part of these interactions reduces the mass at which stars can reach the WR stage \citep{mcclelland_helium_2016}. It is therefore important to constrain the masses of massive binary systems so that we can understand whether or not the WR components were formed as a result of binary interactions. 

Only 3 massive Milky Way binaries with WR stars have masses derived using both spectroscopic and visual binary orbits: $\gamma^2$ Velorum \citep{lamberts_numerical_2017}, WR 133 \citep{richardson_first_2021}, and WR 140 \citep{thomas_orbit_2021}. Fifteen Milky Way WR binary stars have mass estimates derived from spectroscopy, photometry, and polarimetry, although their uncertainties are not well constrained \citep{crowther_physical_2007}. Outside the Milky Way, some progress has been made to determine WR + O masses in the Magellanic Clouds using radiative transfer models of WR and O star spectra \citep{shenar_wolf-rayet_2019, shenar_wolf-rayet_2016}. 

Measuring the masses of non-eclipsing binary stars is a difficult task. The most robust method is to combine spectroscopic radial velocity orbital solutions that provide $M \sin i$ values with astrometric orbits to obtain the inclination angle of the system, and thus derive masses. Obtaining the inclination angle requires alternative solutions in the absence of a visual orbit. Methods that have been used to determine inclination angles without a visual orbit include photometric models that are reliant on wind eclipses \citep[e.g.][]{lamontagne_photometric_1996}, stellar spectral synthesis models \citep[e.g.][]{martins_new_2005}, models of colliding winds \citep[e.g.][]{hill_modelling_2000}, and polarimetric models \citep[e.g.][]{brown_polarisation_1978}. These methods span a wide range of uncertainties, with the most precise being the photometric models \citep[on the order of 50\% formal uncertainty in WR mass,][]{lamontagne_photometric_1996} and the least precise the models based on stellar spectral synthesis. The true uncertainties of the photometric models may be larger \citep[see footnote 11,][]{lamontagne_photometric_1996}. The stellar spectral synthesis method does not report uncertainties, but rather a likely mass range for the O star based on its spectral class, and that informs the inclination angle of the system to determine the WR mass in concert with previous inclination angle estimates \citep[][]{vanbeveren_wr_1998,vanbeveren_evidence_2020}. Furthermore, the methods do not consistently agree on the masses for a given WR + O system, exacerbated by the range of reported uncertainties (or lack thereof). For example, previous estimates for the mass of WR 79 from the photometric and polarimetric methods disagree by more than 3~$\sigma$ \citep{lamontagne_photometric_1996}. This motivates the more robust method that we describe in this paper.

The polarimetic method mentioned previously works by modeling the time-varying linear polarization signal that occurs when the hot winds of the WR star act as electron scattering regions for photons produced both by the WR star and its companion (see Figure~\ref{fig:model_schematic} showing an electron scattering region and WR + O illumination sources). This produces a constant linear polarization from the WR star when the wind is asymmetric \citep{brown_polarisation_1977}. Illumination by the secondary star produces the time-varying linear polarization, which is partly dependent on the orbital inclination angle $i$ \citep{brown_polarisation_1978}, and this provides a way to measure inclination angles and thus masses. To recover accurate uncertainties for the inclination angle, we must determine the posterior distribution of the inclination angle through repeated model evaluations. 

There are two primary methods for modeling polarimetric signals: analytic methods and numerical radiative transfer methods. Analytic models suffer from a variety of challenges, including a bias towards high inclination angles and limitations to optically thin scattering regions \citep[$\tau < 0.01$;][]{aspin_polarimetric_1981,brown_effect_1982,simmons_bias_1982,2019PASP..131j8005B,wolinski_confidence_1994}.
In contrast, numerical radiative transfer methods are not biased by inclination angle and can include the effects of high optical depths such as multiple scattering \citep[e.g.,][]{hoffman_effect_2003} though these simulations are computationally expensive. We have used an enhanced version of the Monte Carlo radiative transfer (MCRT) code \textsc{slip} \citep{hoffman_polarized_2007,huk_time-dependent_2017,shrestha_polarization_2018} to perform numerical simulations of binary stars in circular orbits, for cases where one star has a scattering region surrounding it. To obtain sufficient signal-to-noise for numerical polarimetric estimates, which are often much less than 1\% of the total emitted intensity, \textsc{slip} requires large numbers of photons to be propagated through the simulation, slowing its evaluation times. To reconstruct the posterior distribution of the model parameters, we must evaluate the model many times. This motivates the adoption of acceleration methods.

One such method to accelerate models is that of emulators based on neural networks. Emulators relate the approximated model to its input parameters. Neural networks are capable of rapidly approximating any model \citep[see e.g.][]{cybenko_approximation_1989, hornik_multilayer_1989}. In our case, the trained neural network rapidly generates an approximated polarimetric \textsc{slip} model based on the input parameters. This means the emulator can be rapidly sampled to obtain many models and infer parameters using Bayesian statistics with maximum-likelihood testing for problems that lie within the range of the training data. This technique has been successfully applied to spectral models \citep{czekala_constructing_2015, kerzendorf_dalek_2021, 2021ApJ...916L..14O}.


In this paper, we use an enhanced version of the \textsc{slip} radiative transfer code accelerated by an emulator to produce a robust probabilistic estimate of the inclination angles and thus masses of four WR + O binary systems in the Galaxy with existing polarimetric observations. In Section~\ref{sec:observations}, we describe the polarimetric data and the WR + O systems. In Section~\ref{sec:methods}, we describe our radiative transfer model and emulator procedure. In Section~\ref{sec:results}, we present our results and discuss their implications. Appendix~\ref{sec:slip_appdx} describes the SLIP code in detail, and we validate our model in Appendix~\ref{sec:slip_validate}. In Appendix~\ref{sec:emulator_app}, we provide additional details about our emulator method. The full parameter spaces for our investigations of each system are presented in Appendix~\ref{sec:corners}.


\section{Data}\label{sec:observations}

We investigated four WR + O binary systems with publicly available, phased polarization data. We sourced the data from \citet[][WR 42 and WR 79]{st-louis_polarization_1987} and \citet[][WR 127 and WR 153]{st-louis_polarization_1988}. All of the objects were observed with the Minipol polarimeter at Las Campanas, Chile from 1985--1986 \citep{frecker_linear_1976}, using a blue filter with central wavelength 4700\,\AA\ and FWHM 1800\,\AA.  Table~\ref{tab:wr_systems} lists the objects and their estimated orbital characteristics. We refer to the systems via their WR catalogue number as defined in \citet{crowther_wolf-rayet_2015}\footnote{\url{http://pacrowther.staff.shef.ac.uk/WRcat/index.php}}.

\begin{deluxetable*}{llcccccc}
    \tablecaption{\label{tab:wr_systems}WR + O system spectral types and orbital data. Spectral types are taken from \citet{crowther_wolf-rayet_2015}.}
    
    \tablehead{ \colhead{WR} & \colhead{Spectral type}  & \colhead{Period (d)} & \colhead{Source} & \colhead{$i\,(\degr)$} & \colhead{Source} & \colhead{$\Omega\,(\degr)$} & \colhead{Source}}
    \startdata
    42        & WC7 + O7V      & 7.8912     & 1      & 36--44      & 2, 3  & 63.1--72.8    & 6 \\
    79        & WC7 + O5-8     & 8.8911     & 1      & 29--45      & 2, 3  &  103.2--111.2 & 6  \\
    127       & WN5o + O8.5V   & 9.5550     & 2      & 55--90      & 4  &  83.6--103    & 7 \\
    153       & WN6o/CE + O3-6 & 6.6887     & 3      & 65--78      & 5   & 93.5--95.4    & 7 \\
    \enddata
    \tablerefs{1 -- \citet{hill_modelling_2000}, 2 -- \citet{lamontagne_photometric_1996}, 3 -- \citet{hill_modelling_2002}, 4 -- \citet{de_la_chevrotiere_spectroscopic_2011}, 5 -- \citet{demers_quadruple_2002}, 6 -- \citet{st-louis_polarization_1987}, 7 -- \citet{st-louis_polarization_1988}.}
\end{deluxetable*}

These systems have similar orbital periods ($\sim$1 week) and circularized orbits. They are split into two each of WC- and WN-type WR stars with O-type companions. We used the orbital phases calculated by \citet{st-louis_polarization_1987, st-louis_polarization_1988} despite the determination of newer ephemerides \citep[e.g.][]{2021MNRAS.501.4214N}, because any uncertainties in the ephemerides are greatly amplified by the large number of orbits of each system that have occurred since the polarimetric data were taken. The ephemerides in \citet{st-louis_polarization_1987,st-louis_polarization_1988} are closer in date to the observations, and thus the effects of uncertainties on the orbital phase are lowest.

\section{Methods}\label{sec:methods}

For this investigation, we used an enhanced version of the MCRT code \textsc{slip} \citep{hoffman_polarized_2007,huk_time-dependent_2017,shrestha_polarization_2018,shrestha2021polarization} that includes binary star capabilities \citep{fullard_spectropolarimetric_2020}. Appendix~\ref{sec:slip_appdx} gives more information about the design of \textsc{slip}. Using this code, we created a model system and trained a neural network emulator in order to analyze the observations.

\subsection{Model binary system}\label{sec:training_model}

\begin{figure}
\centering
\includegraphics[width=0.8\columnwidth]{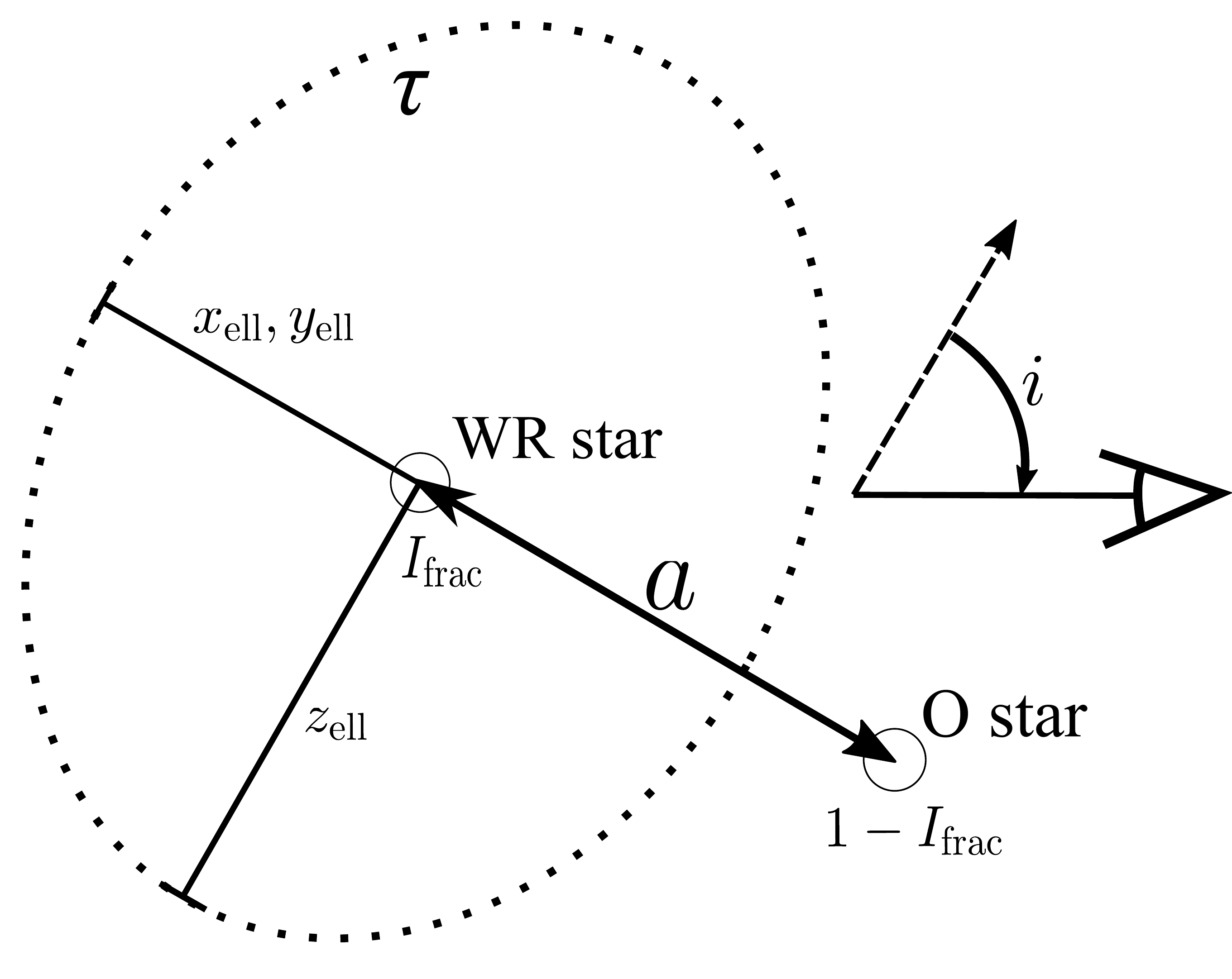}
\caption{Schematic diagram of the binary model. The diagram is presented such that the  system is viewed at an inclination angle of $60\degr$ measured from the normal to the orbital plane. A definition of the model inclination angle $i$ is shown in the bottom right. The stars are the labeled circles of radius $r_\ast$, the scattering region of electron scattering optical depth $\tau$ is the dotted ellipse. The diagram is not to-scale ($r_\ast \ll a$). Other parameters are described in Section~\ref{sec:training_model} and Table~\ref{tab:model_parameters}.}
\label{fig:model_schematic}
\end{figure}

Our model binary system is described using four parameters $I_\mathrm{frac}, \tau, z_\mathrm{ell}$, and $i$. Figure~\ref{fig:model_schematic} shows a schematic diagram of the model. Here, $I_\mathrm{frac}$ is the fractional intensity of the central photon source (the WR star) relative to the total number of photons. Thus, the O star intensity is $1-I_\mathrm{frac}$. $\tau$ is the electron scattering optical depth of the scattering region at $i=90\degr$. 
$z_\mathrm{ell}$ is the half-height of the ellipsoidal scattering region as measured normal to the orbital plane, in arbitrary units. $x_\mathrm{ell}$ and $y_\mathrm{ell}$ are the $x$ and $y$ extent of the scattering region and are held constant and equal to each other. If $z_\mathrm{ell} = x_\mathrm{ell} = 1.5$, the scattering region is spherical. If $z_\mathrm{ell} \neq x_\mathrm{ell}$, the scattering region can be considered prolate or disk-like. $i$ is the inclination (viewing) angle, measured from the normal to the orbital plane. The remaining parameters shown in Figure~\ref{fig:model_schematic} are held constant and are described in Table~\ref{tab:model_parameters}. \textsc{slip} produces simulations of the fractional $q$ and $u$ polarization signal produced by the electron scattering in this model, normalized from the Stokes $Q$ and $U$ vectors as $q=Q/I$ and $u=U/I$.

\begin{deluxetable}{lcl}
\centering
\tablecaption{\label{tab:model_parameters}Constant model parameters for the emulator. Lengths are displayed as relative values (see Fig.~\ref{fig:model_schematic}).}
\tablehead{\colhead{Parameter} & \colhead{Value} & \colhead{Description}}
\startdata
$r_{\ast}$                   & \num{4.65e-4} & Stellar radii       \\
$a$                          & 3.1           & Binary separation            \\
$x_\mathrm{ell}$, $y_\mathrm{ell}$ & 1.5     & $x, y$ extent of the scattering region \\
\enddata
\end{deluxetable}

The $z_\mathrm{ell}$ parameter should not be thought of as the true geometry of the scattering region. Rather, it is a measure of the asymmetry of the scattering region or the density distribution within it (e.g., a disk-like scattering region produces polarization signals similar to a spherical scattering region with equatorially enhanced density). We assume only electron scattering is an important contributor to the polarization signal, which is a reasonable assumption for the case of a highly ionized WR star wind. To reduce the number of parameters required for the model, we assume a constant electron density in the scattering region and a constant albedo of unity, as well as total ionization of the wind. The uniform density means that for scattering regions that are asymmetric, there is a non-linear relationship between $z_\mathrm{ell}$ and $\tau$ as a function of $i$. The constant albedo of one means that the polarization signal is independent of wavelength.

We assume that the stars are sufficiently separated (parameter $a$) that eclipse effects are negligible, which is appropriate for the range of expected inclination angles and periods of our chosen objects. \citet{fox_stellar_1994} showed that occultation is only important in extremely close binary systems, where separation of stars is less than 10 times the radius of the primary. In the case of WR 127, eclipsing inclination angles are unlikely as they would result in a mass for the system that is significantly lower than any other estimate (see comparisons and discussion in Section~\ref{sec:results}). An exploration of eclipse effects would require varying star separation and star size parameters, with a commensurate increase in the number of model runs required to train the emulator.

The sample space of our model, tabulated in Table~\ref{tab:sample_space}, was chosen to represent the four WR + O binaries described in Section~\ref{sec:observations}. The range of scattering region geometries encapsulated by $z_\mathrm{ell}$ transitions from disk-like ($0.2 < z_\mathrm{ell} < 1.5$) through to prolate ($1.5 < z_\mathrm{ell} < 3.1$) distributions. For the WR + O star binaries we have investigated in this paper, the scattering region approximates various geometries (or equivalently, density distributions) of the WR star wind. 
The optical depth range covers scattering from optically thin material (the outer regions of WR star winds) to more optically thick material  \citep[closer to the location of emission lines that drive the wind;][]{sander_driving_2020}. The inclination angle range is valid for Stokes $Q$ and $U$ polarization measurements, which can distinguish between above and below the orbital plane, or equivalently the orbital direction about the orbital plane normal vector.

\begin{deluxetable*}{lDDl}
    \tablecaption{\label{tab:sample_space}Sample space of the model.}
    \tablehead{\colhead{Parameter}     & \twocolhead{Min} & \twocolhead{Max} & \colhead{Description}}
    \decimals
    \startdata
        $I_\mathrm{frac}$ & 0.0   & 1.0  & Fractional intensity of the WR star \\
        $\tau$        & 0.0   & 1.0 & Electron scattering optical depth  \\
        $z_\mathrm{ell}$ & 0.2 & 3.1 & $z$ extent of the scattering region \\
        $i$           & 0.0$\degr$  & 180.0$\degr$ & Orbital inclination angle\\
    \enddata
\end{deluxetable*}

\subsection{Transforming to the observational frame}\label{sec:transformation}

When a binary system is observed in polarized light, two additional factors arise that we do not model using \textsc{slip}. These are interstellar polarization (ISP) and the position angle of the binary system's orbit as projected onto the sky. 

The ISP in the Milky Way at optical wavelengths is described by the empirical Serkowski law \citep{serkowski_wavelength_1975}. We treat it as constant with time for our purposes, because the observations were taken over a timescale that is much shorter than the expected timescale of ISP change \citep[$\sim$30--50 Myr,][]{voshchinnikov_effects_2014}. 

In our case of broadband filter polarimetry, we add ISP to the $q$ and $u$ models separately by calculating the $q_\mathrm{ISP}$ and $u_\mathrm{ISP}$ at the peak wavelength of each filter. This ISP estimate is therefore a flux-weighted ISP across the observed filter.

The observed position angle of the binary orbit manifests itself as a rotation of the observed polarization in the $q$--$u$ plane, and thus applying this rotation to the model moves it to the observational frame. The rotation is applied to the model using

\begin{equation}
    q = p \cos 2 (\theta - \theta_R)
\end{equation}
\begin{equation}
    u = p \sin 2 (\theta - \theta_R)
\end{equation}

\noindent where $p=\sqrt{q_{em}^2+u_{em}^2}$ (total polarization), $\theta$ is the position angle of the  observation, and $\theta_R$ is the rotation angle of the model. Following \citet{villar-sbaffi_first_2005}, we recover the angle of the line of ascending nodes, $\Omega$, using $\Omega = (\theta_R + \pi) / 2$. 

A final correction is needed to match the model's orbitally-phased polarization signal to the observed phases. We project the model onto an orbital phase grid of 40 points, shifted to the center of each orbital phase bin that is output from \textsc{slip}. We linearly interpolate the model onto the observed phases so that the likelihood function (see Section~\ref{sec:likelihood}) is valid for each observed phase $\phi$.

\subsection{Emulator pipeline}\label{sec:emulator}

Typical run times for our model in the sample space are around 2--4 hours (see Appendix~\ref{sec:training_set}). This makes model evaluations intractable for producing probabilistic estimates of parameters. We accelerate the process with an \textit{emulator} which is a machine-learning method, in this case a neural network which is trained to relate model parameters to the output of the model. Such emulators have been successfully applied to astrophysics problems \citep[e.g.\ Type Ia supernovae,][]{2021ApJ...916L..14O}. The emulator learns the relationship between model parameters and output by fitting to a parameter set and its associated model outputs (the \textit{training data}). The user chooses the training data, and these constrain the region over which the emulator is valid.

The construction of an emulator thus requires training data but also a specified neural-network architecture. We trained the emulator on the model described in Section~\ref{sec:training_model}. The emulator training set is described in Appendix~\ref{sec:training_set} and the emulator architecture is described in Appendix~\ref{sec:architecture}. The emulator increases the model evaluation speed by a factor of ${\sim}10^8$ over running \textsc{slip} for a typical model in the sample space in Table~\ref{tab:sample_space}. The emulator representation of \textsc{slip} for the training set is accurate to within 5\% of the \textsc{slip} model.

\subsection{Parameter inference}\label{sec:parameter_inference}

Two components are required for a rigorous parameter inference from our model: a prior distribution and a likelihood function. Obtaining the posterior distribution requires the selection of a prior probability distribution across the parameter space and a likelihood function to compare the data with the model. We describe the prior distributions in Section~\ref{sec:priors}. Polarimetric models are produced by the emulator from samples drawn from the prior distribution.  
We calculate the likelihood of an individual prior input using a $\chi^2$ likelihood function, described in Section~\ref{sec:likelihood}. We describe our exploration of the posterior via Monte Carlo sampling of the priors in Section~\ref{sec:sampling}.

\subsubsection{Priors}\label{sec:priors}

We constructed the Bayesian prior of the parameters listed in Table~\ref{tab:model_parameters} using information about the target object. We sampled the parameters $I_\mathrm{frac}$, $\tau$ and $z_\mathrm{ell}$ uniformly over the entire sample space of the model, because we do not have existing constraints on these parameters that are narrower than the sample space. We chose a normal distribution for the prior of the inclination angle $i$, with the standard deviation set as the half-range of previous inclination angle estimates for each system in Table~\ref{tab:wr_systems}, and the mean taken to be the center of the range. In this way, we incorporate existing information about the system, including photometric and spectroscopic measurements. $I_\mathrm{frac}$ is allowed to vary over the entire emulator sample space. 

For the transformation parameters described in Section~\ref{sec:transformation}, we sample $q_\mathrm{ISP}$ and $u_\mathrm{ISP}$ from Gaussian distributions with the peaks located at the estimated ISP from \citet{fullard_multiwavelength_2020} for each target. We take $\sigma$ values from the propagated uncertainty of $q_\mathrm{ISP}$ and $u_\mathrm{ISP}$, using the uncertainties reported by \citet{fullard_multiwavelength_2020}. For WR 42, we re-fit the ISP following the methods in \citet{fullard_multiwavelength_2020} without the intrinsic constant polarization component, because our model should produce intrinsic constant polarization by varying $I_\mathrm{frac}$, $\tau$ and $z_\mathrm{ell}$. The ISP values are presented in Table~\ref{tab:wr_isp}. The transformation parameter $\theta_R$ is taken as a uniform prior from $-\pi/2\leq\theta_R\leq\pi/2$ because existing estimates for the directly related orbital parameter $\Omega$ were produced with the biased analytic polarization model. Note that this specific range occurs because the polarization position angle is degenerate (repeating in $\pi$ instead of $2\pi$ rotations in the $q$--$u$ plane).

\begin{deluxetable}{lrr}
    \tablecaption{\label{tab:wr_isp}WR + O system ISP estimates calculated from \citet{fullard_multiwavelength_2020} except for WR 42 (see text).}
    \tablehead{\colhead{WR number} & \colhead{$q_\mathrm{ISP}$ (\%)} & \colhead{$u_\mathrm{ISP}$ (\%) }}
    \startdata
    42        & $-0.085\pm0.164$         & $-1.13\pm0.179$  \\
    79        & $-0.284\pm0.020$          & $-0.214\pm0.025$  \\
    127       & \ $0.618\pm0.149$          & \ $0.661\pm0.140$  \\
    153       & $-0.182\pm0.170$          & \ $4.16\pm0.018$  \\     
    \enddata
\end{deluxetable}

\subsubsection{Likelihood}\label{sec:likelihood}

We formulated the likelihood function with a log-likelihood function

\begin{multline}\label{eqn:log_likelihood}
    \log \mathcal{L}(\vec{t}) =\\ -\frac{1}{2}\sum_{\phi}\left[ \left(\frac{q_{em}(\vec{t}) - q_{\rm{obs}}}{\sigma_q}\right)_{\phi}^2 + \left(\frac{u_{em}(\vec{t}) - u_{\rm{obs}}}{\sigma_u}\right)_{\phi}^2\right]
\end{multline}

\noindent where $q_{em}$ and $u_{em}$ are the predicted $q$ and $u$ Stokes vectors from the emulator using parameters $\vec{t}$. $q_{\rm{obs}}$ and $u_{\rm{obs}}$ are the observed Stokes vectors, and $\sigma_q$ and $\sigma_u$ are the associated uncertainties of the observed Stokes vectors. The likelihood function is summed over $\phi$, the orbital phase of the individual $q_{\rm{obs}}$ and $u_{\rm{obs}}$ data points. Stokes $q$ and $u$ are independent, hence we simply sum them as part of the logarithmic transformation of the likelihood function.

The transformation to the observational frame described in Section~\ref{sec:transformation} replaces $q_{em}$ and $u_{em}$ in Equation~\ref{eqn:log_likelihood} with transformed versions. The transformations are applied in the following order: interpolation to the observed phases, then rotation to the observed position angle, then addition of $q_\mathrm{ISP}$ and $u_\mathrm{ISP}$ to $q_{em}$ and $u_{em}$.

\subsubsection{Sampling}\label{sec:sampling}

Following the methodology of \citet{2021ApJ...916L..14O}, we used the UltraNest\footnote{\url{https://johannesbuchner.github.io/UltraNest/}} \citep{2021JOSS....6.3001B} nested sampling package to achieve robust sampling of the posterior distribution with the MLFriends algorithm \citep{buchner_statistical_2016,2019PASP..131j8005B}. The posterior distribution was explored with 800 live points. It converged (based on the UltraNest measures) after $\sim$\num{25000} iterations and required $\sim$\num{6e6} model evaluations for each target. The convergence of the sampling process typically took two hours.

\section{Results and discussion}\label{sec:results}

\begin{figure*}
\centering
\includegraphics[width=\textwidth]{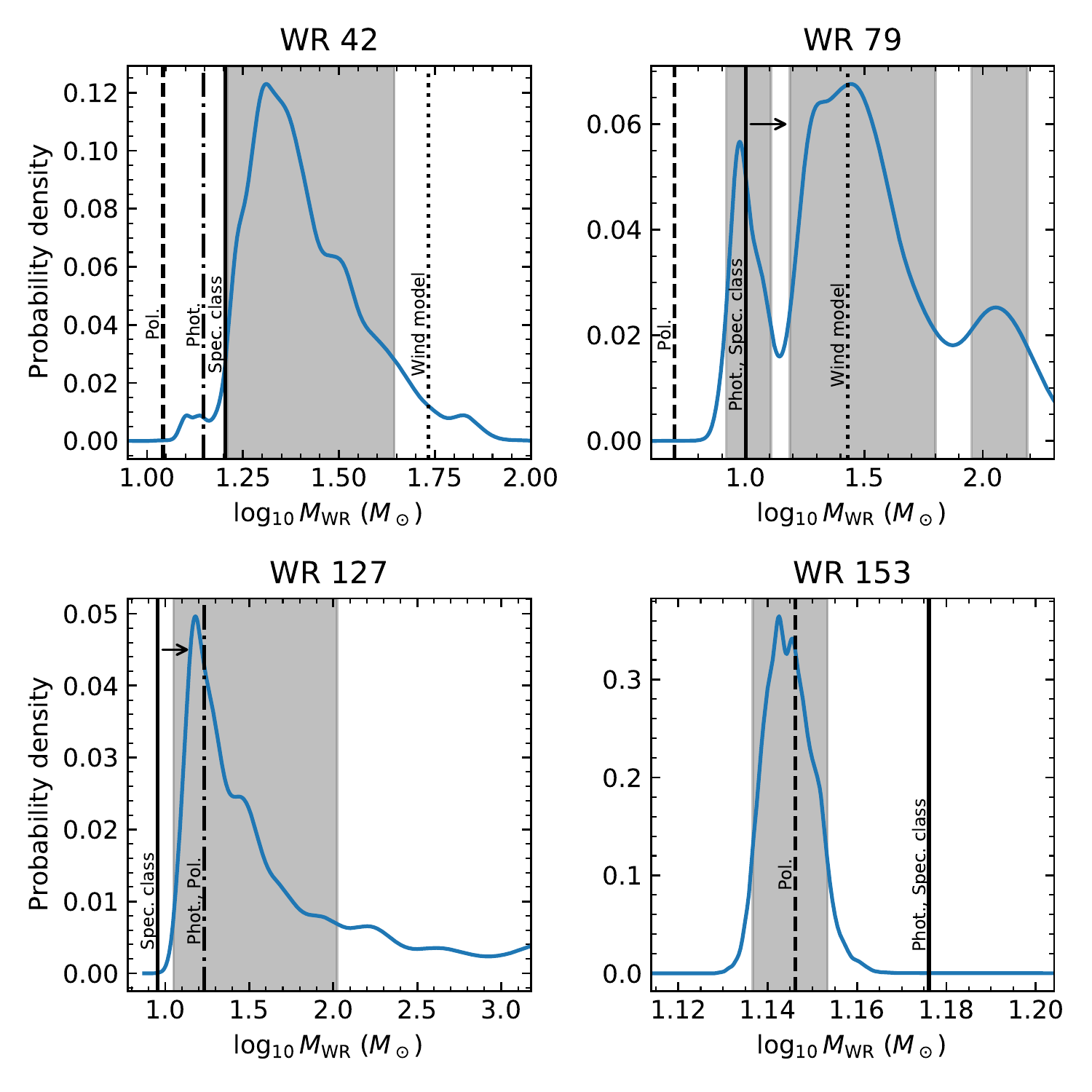}
\caption{WR 42, WR 79, WR 127, WR 153 WR star mass posterior distributions shown as a solid curve. 
The grey shaded region shows the 90\% credible interval of the distribution calculated using the highest density interval algorithm. Vertical lines show previous mass estimates for the systems and are labelled with the method that was used to obtain the estimate. Spec. class: spectral class mass estimates from \citet[][WR 42, WR 79, WR 127]{vanbeveren_evidence_2020} and \citet[][WR 153]{shao_nonconservative_2016}. Arrows show lower limits. Phot: photometric mass estimates from \citet{lamontagne_photometric_1996}. Pol: polarimetric estimates from \citet[][WR 42, WR 79]{st-louis_polarization_1987} and \citet[][WR 127, WR 153]{st-louis_polarization_1988}. Wind model: mass estimates using a spectroscopic colliding wind model from \citet{hill_modelling_2000}.
}
\label{fig:inclination_masses}
\end{figure*}

We sampled the posterior distribution following the prescription described in Section~\ref{sec:sampling}, which produced a 6-dimensional posterior distribution across the parameter space (one dimension for each parameter). Figure~\ref{fig:inclination_masses} shows the posterior distributions for the masses of the WR star components, derived from the inclination angles using $M_{WR}\sin^3 i$ measurements taken from \citet{davis_orbit_1981}, \citet{seggewiss_investigation_1974}, \citet{de_la_chevrotiere_spectroscopic_2011}, and \citet{lamontagne_photometric_1996}. Existing mass estimates for the systems we investigated are based on spectral class \citep{shao_nonconservative_2016, vanbeveren_evidence_2020}, analysis of their polarimetric behaviour \citep{st-louis_polarization_1987, st-louis_polarization_1988}, colliding wind models \citep{hill_modelling_2000}, and photometric eclipse models \citep{lamontagne_photometric_1996}. Of these existing mass estimates, the spectral class models do not provide uncertainties.  

\begin{figure*}
\centering
\includegraphics[width=\textwidth]{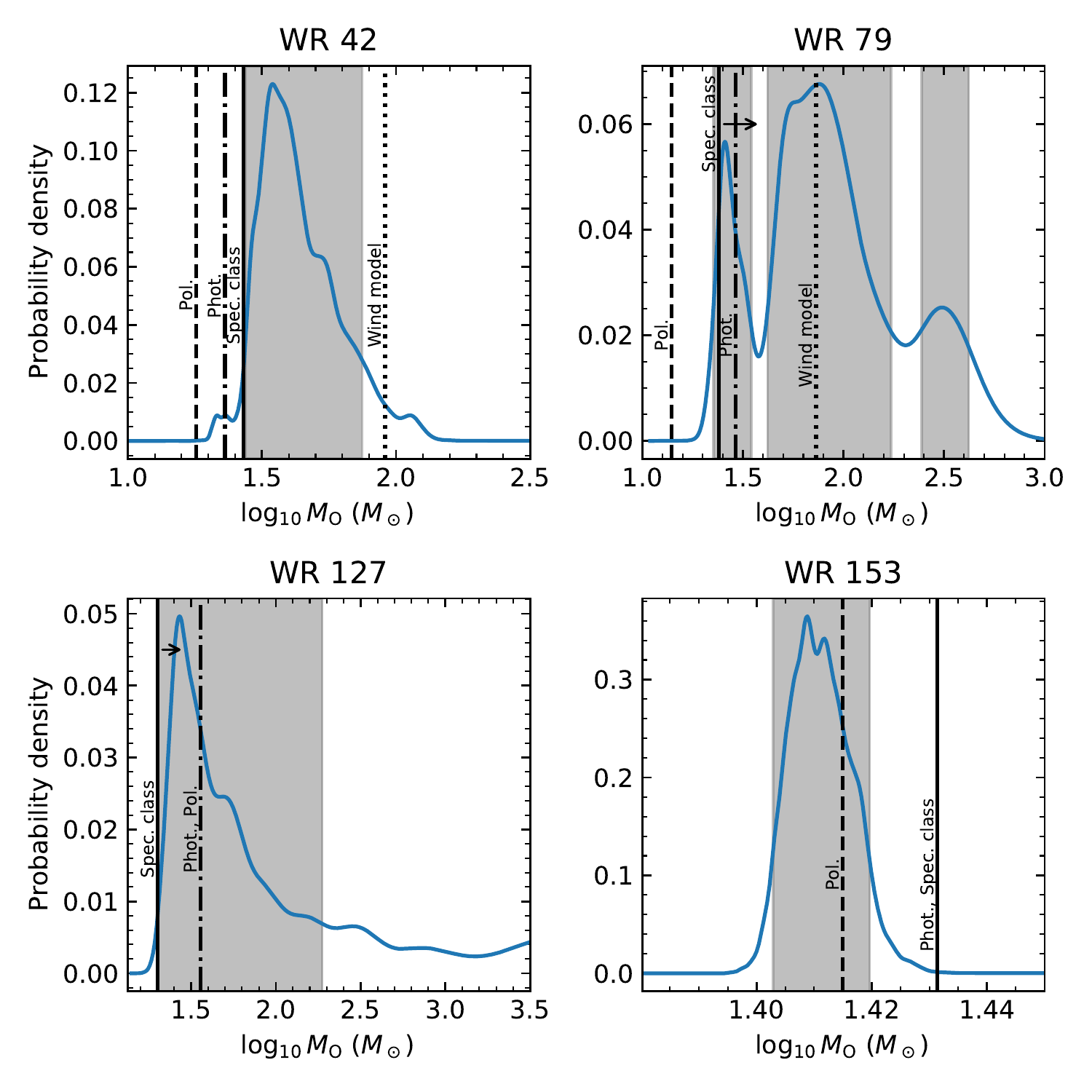}
\caption{WR 42, WR 79, WR 127, WR 153 O star mass posterior distributions shown as a solid curve. 
The grey shaded region shows the 90\% credible interval of the distribution. Vertical lines show previous mass estimates for the systems and are labelled with the method that was used to obtain the estimate (see the caption to Fig.~\ref{fig:inclination_masses}). 
}
\label{fig:inclination_masses_o}
\end{figure*}

Figure~\ref{fig:inclination_masses_o} shows the posterior distributions of masses for the O star components. Table~\ref{tab:wr_masses} shows the 90\% credible interval(s) for the component masses of each object, and we discuss these results below. The credible intervals were calculated using the highest density interval algorithm of the \textsc{arviz} Python package \citep{kumar_arviz_2019}. We display the full set of posterior distributions as corner plots in Appendix~\ref{sec:corners}.


\begin{deluxetable}{lr@{--}lr@{--}lc}
\tablecaption{\label{tab:wr_masses}Our new WR + O system mass estimates derived from our radiative transfer models, including all modes for WR 79.}
\tablehead{
\colhead{WR number} & 
\twocolhead{$M_\mathrm{WR}$ (M$_\odot$)} & 
\twocolhead{$M_\mathrm{O}$ (M$_\odot$)} & 
\colhead{$M\sin^3 i$ source}
}
    \startdata
    42          &   16.2 & 44.0    &  27.4 & 74.6     & 1\\
    79 (mode 1) &    8.3 & 12.8    &  22.5 & 34.8     & 2\\
    79 (mode 2) &   15.3 & 62.8    &  41.8 & 170.9    & 2\\
    79 (mode 3) &   89.5 & 152.2   & 243.7 & 414.2    & 2\\
    127         &   11.1 & 102.5   &  19.8 & 183.1    & 3\\
    153         &   13.7 & 14.2    &  25.3 & 26.3     & 4\\
    \enddata
\tablerefs{1 -- \citet{davis_orbit_1981}, 2 -- \citet{seggewiss_investigation_1974}, 3 -- \citet{de_la_chevrotiere_spectroscopic_2011}, 4 -- \citet{lamontagne_photometric_1996}}
\end{deluxetable}

\subsection{WR 42}\label{sec:wr42}

The WR mass 90\% credible interval calculated from our method for WR 42 (16.2--44.0 M$_\odot$) agrees with existing mass estimates from all of the methods displayed in Figure~\ref{fig:inclination_masses}, taking into account their uncertainties. For the previous polarimetric analysis \citep{st-louis_polarization_1987}, the estimate was 8--14 M$_\odot$, however this did not fully take into account the bias in the analytical polarimetric model \citep{wolinski_confidence_1994} and thus the true uncertainty is larger. The photometric method mass estimate \citep{lamontagne_photometric_1996} did not provide WR mass uncertainties, though the reported inclination angle of 40.3\degr$\pm$2.9 produces a WR mass range of 11--16 M$_\odot$. It is important to note that the inclination angle uncertainty here is the ``formal'' uncertainty and the real values can be larger \citep[see footnote 11, ][]{lamontagne_photometric_1996}. The wind model of \citet{hill_modelling_2000} produces a very uncertain mass estimate of 54$\pm$43 M$_\odot$ so we provide it for the sake of completeness. Our derived 90\% credible interval for the O star mass agrees with the existing mass estimates from spectral class models, encapsulating the \citet{vanbeveren_evidence_2020} estimated mass range of 27--37 M$_\odot$. The WR mass reported by \citet{vanbeveren_evidence_2020} is ``adopted'' as 16 M$_\odot$, based on the previous reported inclination angles for the system (and thus similarly uncertain as the polarimetric and photometric mass estimates).

It is interesting to note that the polarimetric measurement from \citet{st-louis_polarization_1987} produced using the same data overlaps (within the \citet{st-louis_polarization_1987} polarimetric uncertainties) a small secondary peak in our posterior distribution. This may indicate a local minimum was found with the analytic model. If the uncertainty in $M \sin^3 i$ reported by \citet{davis_orbit_1981} for WR 42 is used to calculate the mass range, this extends both the WR and O star mass 90\% credible intervals to 25--48 M$_\odot$ and 19--81 M$_\odot$, respectively. 

\citet{fullard_multiwavelength_2020} reported that the mean polarization signal of WR 42 had a large constant intrinsic component in both $q$ and $u$. As noted in Section~\ref{sec:priors}, we fit the ISP without the intrinsic polarization terms. As a result, the model predicts a strongly asymmetric WR wind that is more disk-like, or equivalently a symmetric WR wind with a strong density increase parallel to the orbital plane. This may explain the intrinsic polarization, and potentially represents the asymmetry produced by the colliding wind structures proposed by \citet{hill_modelling_2000}. 

\subsection{WR 79}

We find multiple peaks in the WR 79 mass posterior distribution. This creates a series of three modes with associated credible intervals, which we report from low to high mass in Table~\ref{tab:wr_masses}. The first credible interval (8.3--12.8 M$_\odot$) agrees with both the spectral class model mass lower limit of 10 M$_\odot$ \citep{vanbeveren_evidence_2020} and the photometric model mass estimate of 9--13 M$_\odot$ \citep{lamontagne_photometric_1996} The second, more mathematically probable posterior peak and associated credible interval (15.3--62.8 M$_\odot$) matches the mass estimate from the colliding wind model of \citet{hill_modelling_2000}. However, we note that the wind model mass estimate is very uncertain at 27$\pm$17 M$_\odot$ \citep{hill_modelling_2000} and therefore it also overlaps with the first credible interval. The third credible interval (89.5--152.2 M$_\odot$) is very unlikely, and suggests masses well outside any previous model. None of the intervals agree with the previous polarimetric mass estimate that used the same data \citep{st-louis_polarization_1987}, confirming that the basic analytic model is unlikely to provide accurate results for binary systems with low inclination angles. Note that our model recovers an inclination angle greater than 90$\degr$, which is possible for polarimetry because the sign of Stokes $U$ breaks the degeneracy present in spectroscopic orbital models. This inclination angle is equivalent to a reversed orbital motion compared to the model's motion in the inclination range $0\degr\leq i \leq 90\degr$.

\begin{figure}
\gridline{\fig{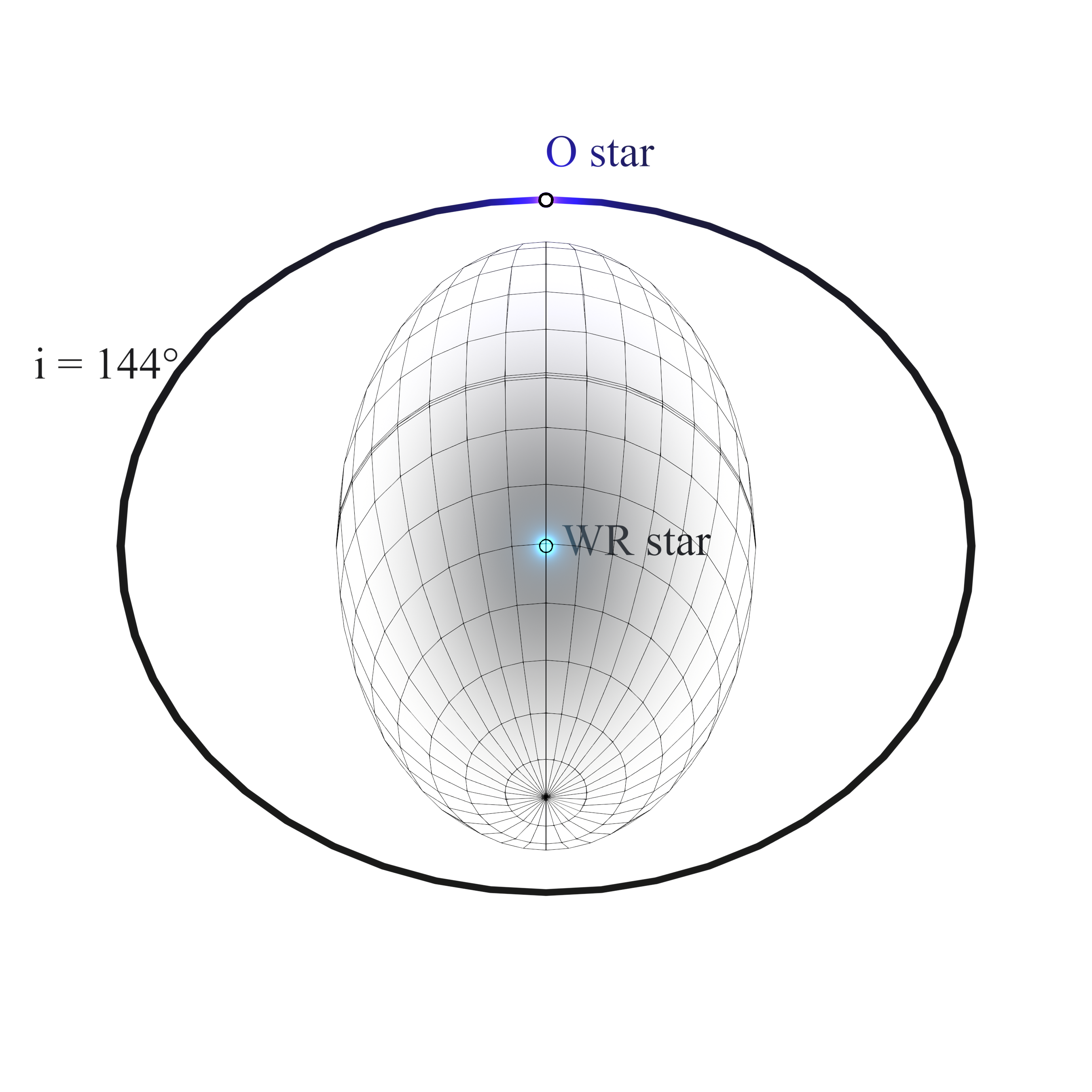}{\columnwidth}{(a)}}
\gridline{\fig{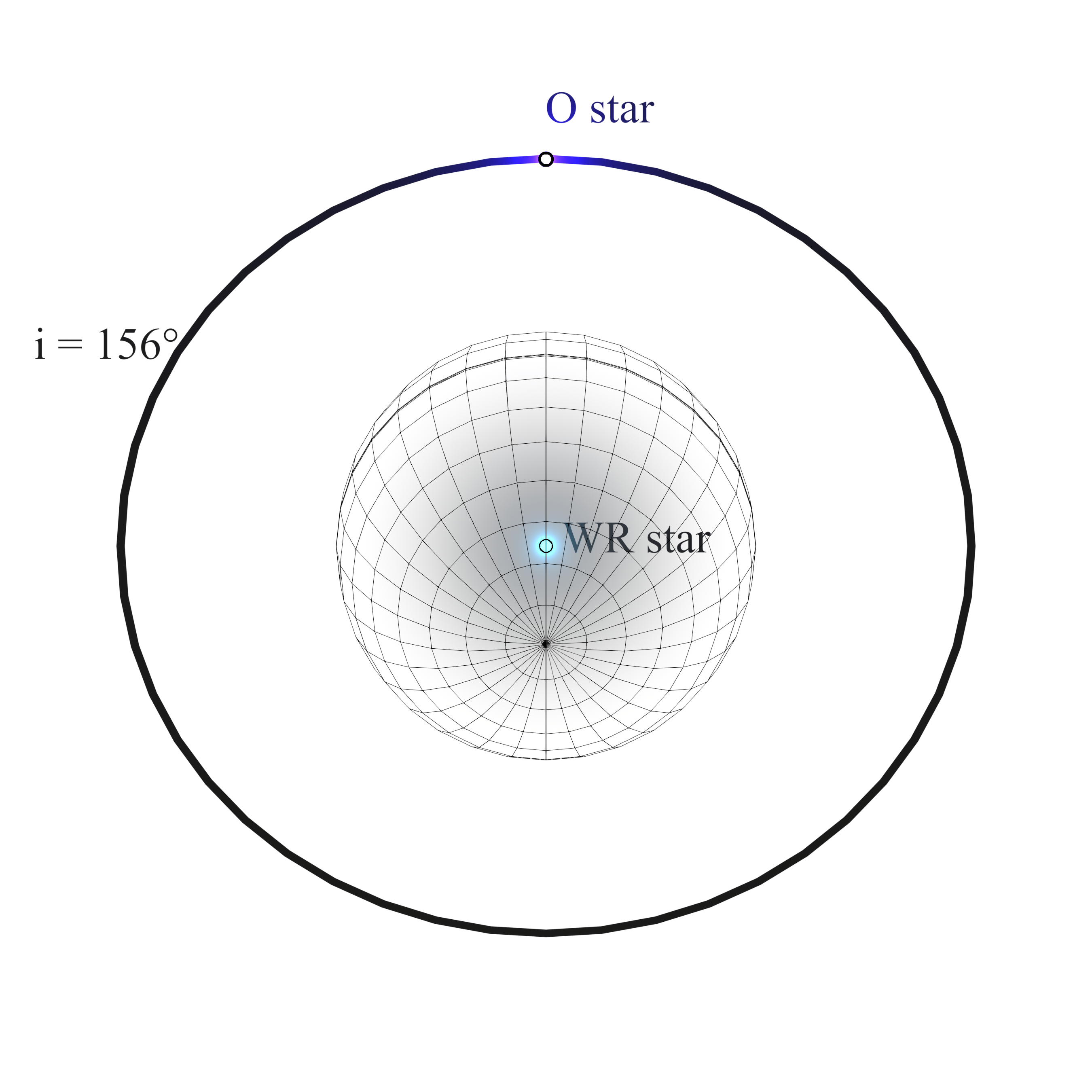}{\columnwidth}{(b)}}
\caption{\textbf{(a)} WR 79 representative model of the first posterior mode of mass 8.3--12.8 M$_\odot$ at $i=144\degr$. \textbf{(b)} WR 79 representative model of the second posterior mode of mass 15.3--62.8 M$_\odot$ at $i=156\degr$. Both figures are at orbital phase 0.5 and to scale. The triple line on the scattering regions shows the orbital plane, and the semitransparent fill represents their optical depths. The thick black line is the projected O star orbit. The stars have relative emission equivalent to each model}
\label{fig:wr79_models}
\end{figure}

The physical implications of the multiple modes are of some interest. Scale renderings of the models are shown in Figure~\ref{fig:wr79_models}. The first inclination angle (mass) mode (top in Fig.~\ref{fig:wr79_models}) corresponds to models with an extremely asymmetric WR wind, low electron scattering optical depth of $\tau<0.1$, and the majority of emission ($\sim95\%$) originating from the O star. The asymmetric wind can also be considered to be a strong polar density enhancement. However, such an enhancement is unlikely. Distortions to the WR wind caused by wind-wind collisions are observed to occur primarily in the plane of the orbit \citep[e.g.][]{callingham_two_2020} and models agree \citep[e.g.][]{lamberts_numerical_2017, hill_modelling_2000}. Corotating interaction regions \citep[see e.g.][]{mullan_corotating_1984} are another potential source of wind asymmetries, but their effect on polarization is small unless they have an optical depth of order unity \citep{carlos-leblanc_monte_2019}, which would disagree with this first mass mode.

The second mass mode (bottom in Fig.~\ref{fig:wr79_models}) corresponds to models with a nearly spherical WR wind (with $z_\mathrm{ell}>x_\mathrm{ell}$, or equivalently a slight polar density enhancement), an electron scattering optical depth of $\tau\sim0.3$, and the majority of emission ($\sim70\%$) originating from the WR star. The second WR mass mode at $\sim 27$ M$_\odot$ would further increase the similarities between WR 42 and WR 79, which are identical in WR type, similar in O star type, and with orbital periods only one day apart. The corresponding second mode O star mass of $\sim 60$ M$_\odot$ is compatible with the upper estimate for the O star from \citet{vanbeveren_evidence_2020}. Another implication of the higher WR star mass is that the WR star would be more likely to have reached the WR stage without interacting with its companion, based on single-star evolutionary tracks \citep[][]{sander_galactic_2019}.

\subsection{WR 127}

The posterior for the mass of the WR component of WR 127 (90\% credible interval 11.1--102.5 M$_\odot$) is in agreement with \citet{vanbeveren_evidence_2020}, since they provide only a lower limit. We also agree with both photometric and polarimetric estimates. However, our method provides an upper limit on the mass of 49.5 M$_\odot$. It also increases the mass of the lower limit to 11.1 M$_\odot$. The most probable mass is $\sim$14 M$_\odot$. The same is true for the O star, with its most probable mass near $\sim$25 M$_\odot$, at the upper end of the 17--24 M$_\odot$ estimate given by \citet{vanbeveren_evidence_2020}. 

Our mass estimate also agrees with both the previous polarimetric \citep{st-louis_polarization_1988} and photometric \citep{lamontagne_photometric_1996}  methods, which reported WR masses of 11--21 M$_\odot$ and 12--16 M$_\odot$ respectively. Although these are lower uncertainty mass estimates than our presented values, the previous polarimetric estimate was published before the complete bias analysis of \citet{wolinski_confidence_1994} and thus the uncertainty is likely to be larger in reality. Furthermore, the photometric model assumes a spherical WR wind and formal uncertainties. This is unlikely for WR 127 given the evidence for a possible wind collision from X-rays \citep[see Section~\ref{sec:asymmetry} and ][]{2021MNRAS.501.4214N}.

\subsection{WR 153}\label{sec:wr153}

We find that WR 153 has a significantly lower mass (90\% credible interval 13.7--14.2 M$_\odot$) than measured by photometric and spectral class models (where we define ``significant'' as outside the 90\% credible interval), with a low uncertainty compared to the other stars in the sample. The photometric model inclination angle estimate is similarly certain, with a mass range of 14.5--14.7 M$_\odot$ \citep{lamontagne_photometric_1996}, though the true uncertainty is likely to be larger. Similarly to \citet{vanbeveren_evidence_2020}, \citet{shao_nonconservative_2016} simply present an adopted mass for the WR star of 15 M$_\odot$ with no associated uncertainty. Our posterior distribution agrees with the analytic polarimetric estimate using the same data \citep{st-louis_polarization_1988}, which is unsurprising, given that the high inclination angle of the system reduces the bias of the analytic model. The O star mass is low for its spectral type, which is expected to be greater than 30 M$_\odot$ \citep{vanbeveren_wr_1998}. However, there is not currently a measure of its luminosity class, which makes this expectation very uncertain. Furthermore, measurement uncertainties for $M \sin^3 i$ would broaden the range of masses, potentially including the mass estimate from \citet{shao_nonconservative_2016}; \citet{demers_quadruple_2002} produced an additional measurement of $M \sin^3 i$, but it was only a lower limit. 

\subsection{Wind asymmetry}\label{sec:asymmetry}

Our findings indicate that all four systems we investigated have some form of asymmetry in their winds. It is unsurprising to find asymmetry in the winds of close massive stellar binaries because of the possibility of wind--wind interaction between the components. \citet{hill_modelling_2000} modeled the colliding winds in both WR 42 and WR 79, and found that optical line variations could be explained by the presence of conic wind collision regions, making it plausible that our detection of wind asymmetries in these systems is caused by wind-wind collision. Recent work by \citet{2021MNRAS.501.4214N} has shown that WR 127 exhibits X-ray emission that indicates the presence of wind collision regions, though WR 42 was a non-detection and WR 153 showed only faint X-rays compared to the expectation for a wind collision. Thus, the possible wind asymmetry we find for WR 153 may not be caused by wind-wind collision. These results also cast some doubt on the presence of wind--wind collision regions in WR 42. Instead, the wind asymmetries could be caused by rotation of the WR star expanding the wind equatorially \citep{vink_wolf-rayet_2017} or optically thick corotating interaction regions producing areas of enhanced density in the wind \citep{carlos-leblanc_monte_2019}.



\subsection{Other model parameters}\label{sec:other_parameters}
Our posterior distribution reveals relationships between parameters.
Across all models, we note relationships between $\tau$--$z_\mathrm{ell}$, $I_\mathrm{frac}$--$\tau$, $I_\mathrm{frac}$--$z_\mathrm{ell}$, and $q_\mathrm{ISP}$--$u_\mathrm{ISP}$. In the case of $\tau$--$z_\mathrm{ell}$ and $I_\mathrm{frac}$--$\tau$, the two-dimensional posterior distributions show strong exponential-like relationships. The $I_\mathrm{frac}$--$z_\mathrm{ell}$ correlation is similar, but generally less strong. This is unsurprising, given that the probability of a photon scattering in a path length $l$ is $1-\exp(-\tau)$, and $\tau \propto l$. $z_\mathrm{ell}$ changes the path length and thus the required optical depth to reach a given polarization. Therefore, $z_\mathrm{ell}$ and $\tau$ are directly related. Then, a more asymmetric scattering region produces more polarization for a given $I_\mathrm{frac}$, which produces the $I_\mathrm{frac}$--$z_\mathrm{ell}$ relationship; the relationship between $\tau$--$z_\mathrm{ell}$ follows from $I_\mathrm{frac}$--$z_\mathrm{ell}$ and $\tau$--$z_\mathrm{ell}$. Finally, the relationship $q_\mathrm{ISP}$--$u_\mathrm{ISP}$ occurs because the rotation of the model by $\theta_R$ transforms the polarization directly between $q$ and $u$.

As part of the modeling process, we also recover ISP and $\Omega$ posterior distributions. Our ISP posteriors are strongly influenced by the priors, though all the systems show some deviation from the existing estimates in \citet{fullard_multiwavelength_2020}. We present the maximum likelihood estimates in Table~\ref{tab:wr_isp_new}. The uncertainties for the distributions are displayed in Appendix~\ref{sec:corners} as part of the corner plots, split into upper and lower 34\% quantiles.  

\begin{deluxetable}{lDD}
    \tablecaption{\label{tab:wr_isp_new}WR + O system ISP maximum likelihood estimates from our analysis. Uncertainties are displayed in Appendix~\ref{sec:corners} as part of the corner plots.}
    \tablehead{\colhead{WR number} & \twocolhead{$q_\mathrm{ISP}$ (\%)} & \twocolhead{$u_\mathrm{ISP}$ (\%)}}
    \decimals
    \startdata
    42        & -0.52         & -0.49  \\
    79        & -0.31          & -0.24  \\
    127       & 0.69         & 0.68  \\
    153       & -0.04          & 4.15  \\     
    \enddata
\end{deluxetable}

\begin{figure}
\centering
\includegraphics[width=\columnwidth]{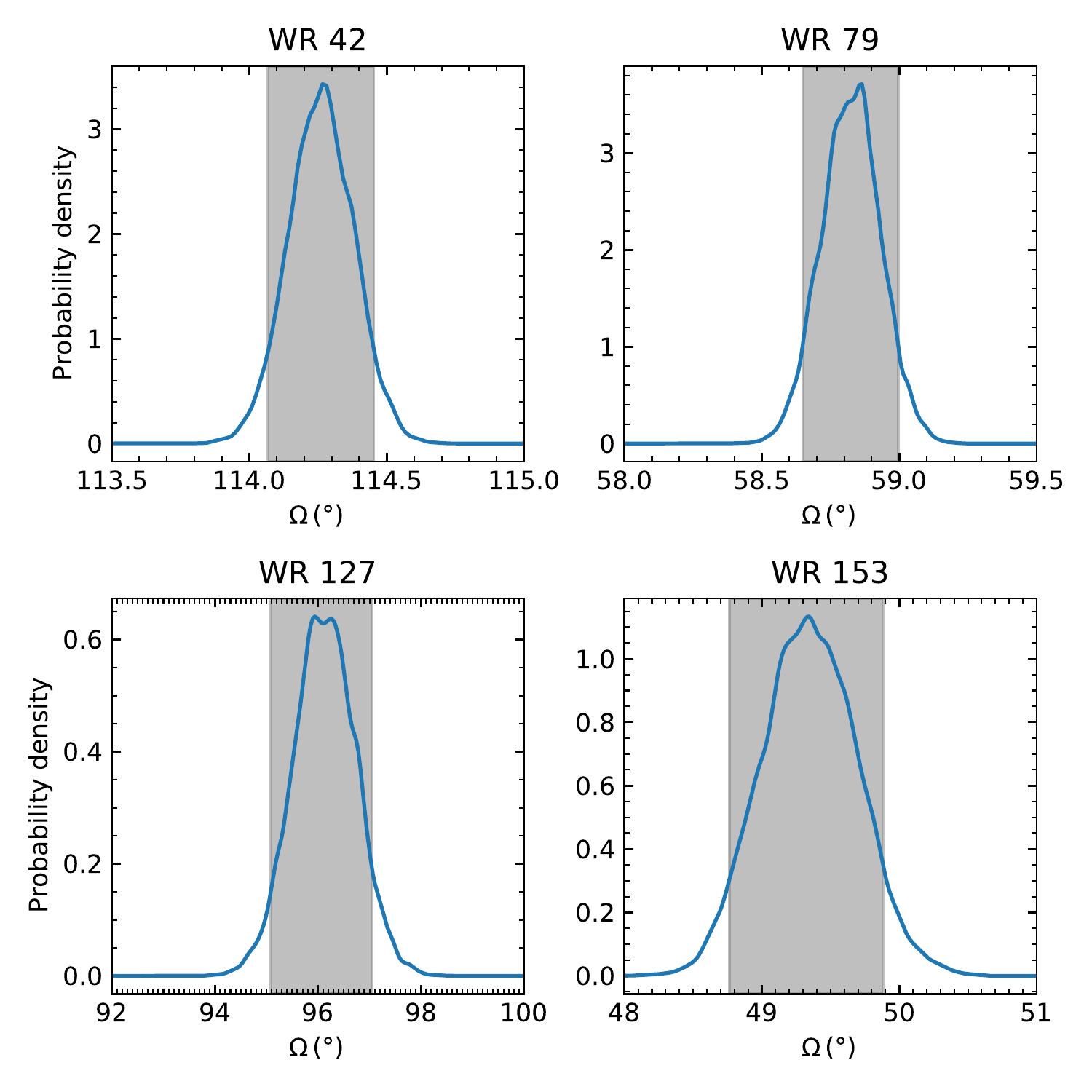}
\caption{Posterior distributions of the longitude of the ascending node, $\Omega$, for WR 42, WR 79, WR 127, and WR 153 shown as solid lines. The shaded area in each panel shows the 90\% credible interval.}
\label{fig:omega}
\end{figure}

Figure~\ref{fig:omega} shows the posterior distribution of $\Omega$ for each system. The grey shaded region shows the 90\% credible interval of the distribution. Our posterior distributions for $\Omega$ are in complete disagreement with the previous estimates presented in Table~\ref{tab:wr_systems} for all but WR 127 (i.e., the previous estimates lie  outside the plotted range of $\Omega$ in Figure~\ref{fig:omega}). However, this is not surprising given that the sources of these estimates did not provide uncertainty estimates for $\Omega$ that take into account the bias on the analytic model that was used to derive $\Omega$ \citep{wolinski_confidence_1994}.

The final parameters we recover are $I_\mathrm{frac}$ and $\tau$, the fraction of photons arising from the WR star and the optical depth of the scattering region, respectively. Our method confirms that these parameters are related to each other and to $z_\mathrm{ell}$. This means that the posterior distributions are difficult to interpret as standalone results. For example, in the case of WR 79, the $I_\mathrm{frac}$ distribution peaks at $z_\mathrm{ell}\sim1.6$ and $\tau\sim0.2$. There are also peaks in the posteriors that correspond to values of $\tau\sim0$, $z_\mathrm{ell}\sim3.0$, $I_\mathrm{frac}\sim0$. Although these two possible models are peaks in the posterior distributions, there is a range of possible models between these peaks because of the relationship between the parameters. If we disregard these relationships and focus on the median $\tau$ for each object, we typically find low optical depths, indicating that the broadband polarization arises in the outer regions of the WR star wind.





\section{Conclusions}\label{sec:conclusion}

We have found probabilistic estimates for the mass of four WR + O binary systems, with a new method based on an emulated radiative transfer model. Our results are different to previous estimates for WR 153, and we provide robust upper limits in all cases, as well as information about degeneracies in the model. Our disagreement with existing polarimetric estimates for these systems shows that care must be taken when drawing conclusions using masses derived from analytic polarimetric models.

We have shown that our method can break the orbital inclination degeneracy in $M \sin^3 i$, but this still requires spectroscopic determination. Thus, we recommend improvements to spectroscopic measurements of $M \sin^3 i$ to obtain more certain mass estimates for the four systems presented in this paper, and others like them. Phased polarimetric measurements are also a requirement for this method, and thus more accurate polarimetric observations are necessary to apply this method to more targets. We are obtaining accurate phased polarimetric observations of a larger sample of Milky Way WR + O systems \citep{fullard2018specpol,johnson2019comparison}, and will apply this method to them in the future. In principle, the model can also be applied to the 30\% of Be stars that have companions \citep{oudmaijer_binary_2010}, because their disks produce polarization signals. Further improvements to our model could be made with the incorporation of elliptical orbits, to investigate systems such as WR 133 that already have polarimetric observations available.

With the advent of telescopes like the Polstar UV spectropolarimetric satellite \citep{scowen_polstar_2021}, it will be critical to have polarization models that can be explored with statistical approaches to understand the observations. UV observations of WR + O binary systems can provide new constraints for the priors presented here.

\section*{Acknowledgments}

We thank Selma de Mink for her insightful comments about the content of this manuscript.

We gratefully thank Johannes Buchner for UltraNest assistance.

TACC is operated by the Extreme Science and Discovery Environment (XSEDE), which we access under allocation AST-120067 to J. Hoffman.

J. Hoffman is grateful for funding from the National Science Foundation under award AST-1816944, and acknowledges that the University of Denver occupies land within the traditional territories of the Arapaho, Cheyenne, and Ute peoples.

R. Ignace  acknowledges funding support for this research from a grant
with the National Science Foundation, AST-2009412.

\software{\textsc{arviz}, \textsc{jupyter} \citep{jupyter}, \textsc{lmfit}, \textsc{tensorflow}, \textsc{scikit-learn}, \textsc{blender} \citep{blender}, \textsc{corner} \citep{corner}, \textsc{ultranest}}



\bibliographystyle{mnras}
\bibliography{SLIP_refs} 

\section{Contributor Roles}
\begin{itemize}
\item Conceptualization: 
Andrew Fullard,
Wolfgang Kerzendorf
\item Data curation: Andrew Fullard
\item Formal Analysis: 
Andrew Fullard, 
Patrick van der Smagt
\item Investigation: 
Andrew Fullard, 
Patrick van der Smagt
\item Methodology: Andrew Fullard
\item Project administration: Wolfgang Kerzendorf
\item Resources: 
Institute for Cyber-Enabled Research at Michigan State University,
Texas Advanced Computing Center
\item Software: 
Andrew Fullard,
Jennifer Hoffman,
Manisha Shrestha,
Patrick van der Smagt,
John O'Brien
\item Supervision:
Wolfgang Kerzendorf
\item Visualization:
Andrew Fullard
\item Writing – original draft:
Andrew Fullard
\item Writing – review \& editing:
Andrew Fullard,
John O'Brien,
Patrick van der Smagt,
Jennifer Hoffman,
Richard Ignace,
Wolfgang Kerzendorf, 
Manisha Shrestha
\end{itemize}




\appendix

\section{SLIP MCRT code}\label{sec:slip_appdx}

\textsc{slip} is a Fortran + MPI code based on the MCRT method outlined in \citet{whitney_monte_2011}.The simulation grid is a uniformly spaced spherical polar coordinate system in $r$, $\theta$ and $\phi$. Photon packets are emitted from user-specified locations in the grid and propagate through the user-defined scattering region. At each grid cell the optical depth is integrated until the photon scatters or exits the grid cell. The photons are collected as they exit the simulation limits and are binned into different observational directions in $\theta$ and $\phi$, with uncertainties calculated via Poisson statistics. In this way, a single model can be viewed from multiple angles, reducing computation time. For additional details of the \textsc{slip} code, see \citet{huk_time-dependent_2017} and \citet{shrestha_polarization_2018}.

The original \textsc{slip} code included a single photon source at the center of the simulation grid. We used an enhanced version of the code that includes an additional photon source located at an arbitrary distance along the $x$-axis ($\theta=90\degr$, $\phi=0\degr$) from the central source (each source can be a finite size or a point). The number of photons emitted from each source is controlled as a fraction of the total photon count. We simulate circular orbits by simply moving the ``observer'' around the grid in the $\phi$ direction. In Appendix~\ref{sec:slip_validate}, we provide validation of these new capabilities against the existing analytic model of \citet[][]{brown_polarisation_1978}, while being aware of the inherent limitations described in Appendix~\ref{sec:slip_validate}.

\section{Validating the SLIP binary model}\label{sec:slip_validate}

To test the validity of the enhanced \textsc{slip} code, we computed models with three circumstellar geometries that can be fit with the analytic model of \citet{brown_polarisation_1978}; that is, a spherical distribution of CSM, a circular disk-like distribution in the orbital plane with an opening angle of $1.8\degr$ (a single $\theta$ angle bin width in height), and a prolate distribution. The prolate distribution is described by an ellipsoid with the major axis parallel to the $z$-axis, described by

\begin{equation}\label{eqn:prolate}
  \frac{x^2}{0.2^2} + \frac{y^2}{0.2^2} + \frac{z^2}{0.6^2}=1  
\end{equation}

\noindent where $x$, $y$ and $z$ refer to the standard Cartesian coordinate system that \textsc{slip} transforms to its spherical coordinate system. The denominators are the values of $x_\mathrm{ell}$, $y_\mathrm{ell}$ and $z_\mathrm{ell}$, the $x$, $y$ and $z$ axes of the ellipsoid.

The finite nature of the \textsc{slip} grid means that the disk-like distribution is a single grid cell thick in the $\theta$ direction. All three geometries were centered on the source located at the origin. Both stellar sources emitted an equal proportion of the photons, and both were set as point sources so eclipse effects are not important and photons are not absorbed within either star. We set the optical depth of all these CSM distributions at $\theta=90\degr$ to $\tau=0.1$ for $0<\phi<2\pi$ to match the optically thin assumption of the analytic model (\citealt[][]{carlos-leblanc_monte_2019} found that this was the limit between optically thin and optically thick for their similar MCRT models). Note that the CSM density is set to be constant and proportional to the optical depth at $\theta=90\degr$, so the prolate model shows an increase in optical depth towards $\theta=0\degr$ and $180\degr$ because of its increased radius towards those angles. Model calculations were made on the Stampede high-performance computing cluster at the Texas Advanced Computing Center\footnote{\url{https://www.tacc.utexas.edu/systems/stampede}}, with $8\times10^9$ photons per simulation. The output was binned to 23 inclination bins and 40 orbital phase bins. 

We fit the analytic model to the numerical results using \textsc{lmfit} \citep{newville_lmfit:_2014} to minimize $\chi^2$, and algebraically calculated the orbital parameters $i$ and $\Omega$ from the Fourier coefficients following \citet{drissen_polarimetric_1986}. The Fourier coefficients $q_i$ and $u_i$ are found by fitting the following equations

\begin{equation}
    q = q_0 + q_1\cos\phi + q_2\sin\phi + q_3\cos2\phi + q_4\sin2\phi
\end{equation}
\begin{equation}
    u = u_0 + u_1\cos\phi + u_2\sin\phi + u_3\cos2\phi + u_4\sin2\phi,
\end{equation}

\noindent where $\phi$ is the azimuthal coordinate and $\phi/2\pi$ is the equivalent orbital phase. The fits had median $\chi_{\nu}^2$ values of 3.30, 2.85, and 0.90 for the disk-like, spherical, and prolate density distributions respectively (using $\chi_{\nu}^2 = \chi^2 / (N - 10)$, where $N$ is the number of data points and 10 is the number of fit variables). The $\chi_\nu^2$ value of 0.9 suggests some over-fitting or overestimate of uncertainties in the case of the disk-like density distribution.

\begin{figure}
\centering
\includegraphics[width=0.85\columnwidth]{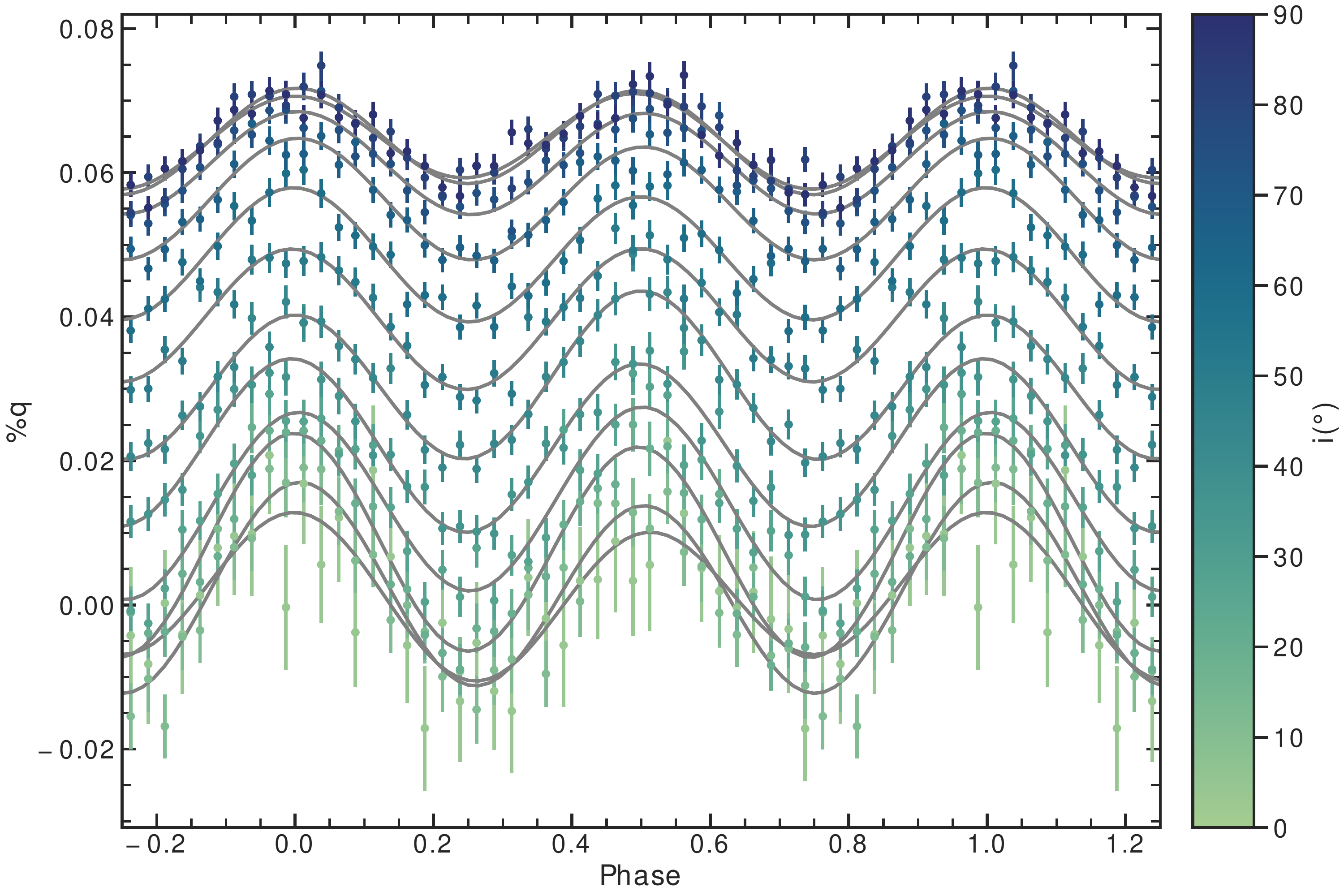}
\includegraphics[width=0.85\columnwidth]{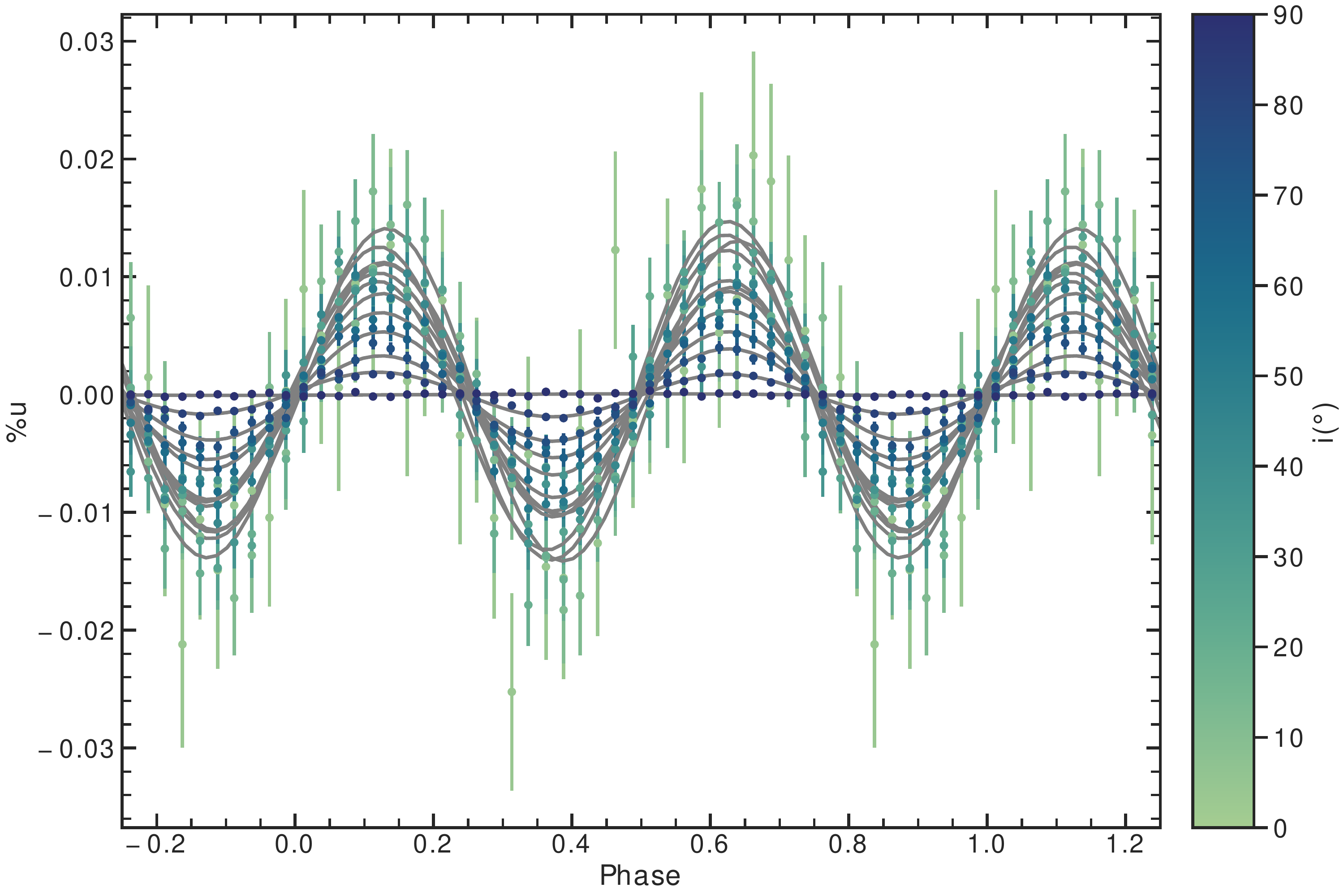}
\caption{Comparison of \textsc{slip} results for a thin disk to the analytic model of \citet{brown_polarisation_1978}. The circles represent \textsc{slip} results for a range of inclination angles from $0-90\degr$ as displayed in the colour bar. The data have been fit by the model of \citet{brown_polarisation_1978} (grey lines).}
\label{fig:bme_disk_slip}
\end{figure}

\begin{figure}
\centering
\includegraphics[width=0.85\columnwidth]{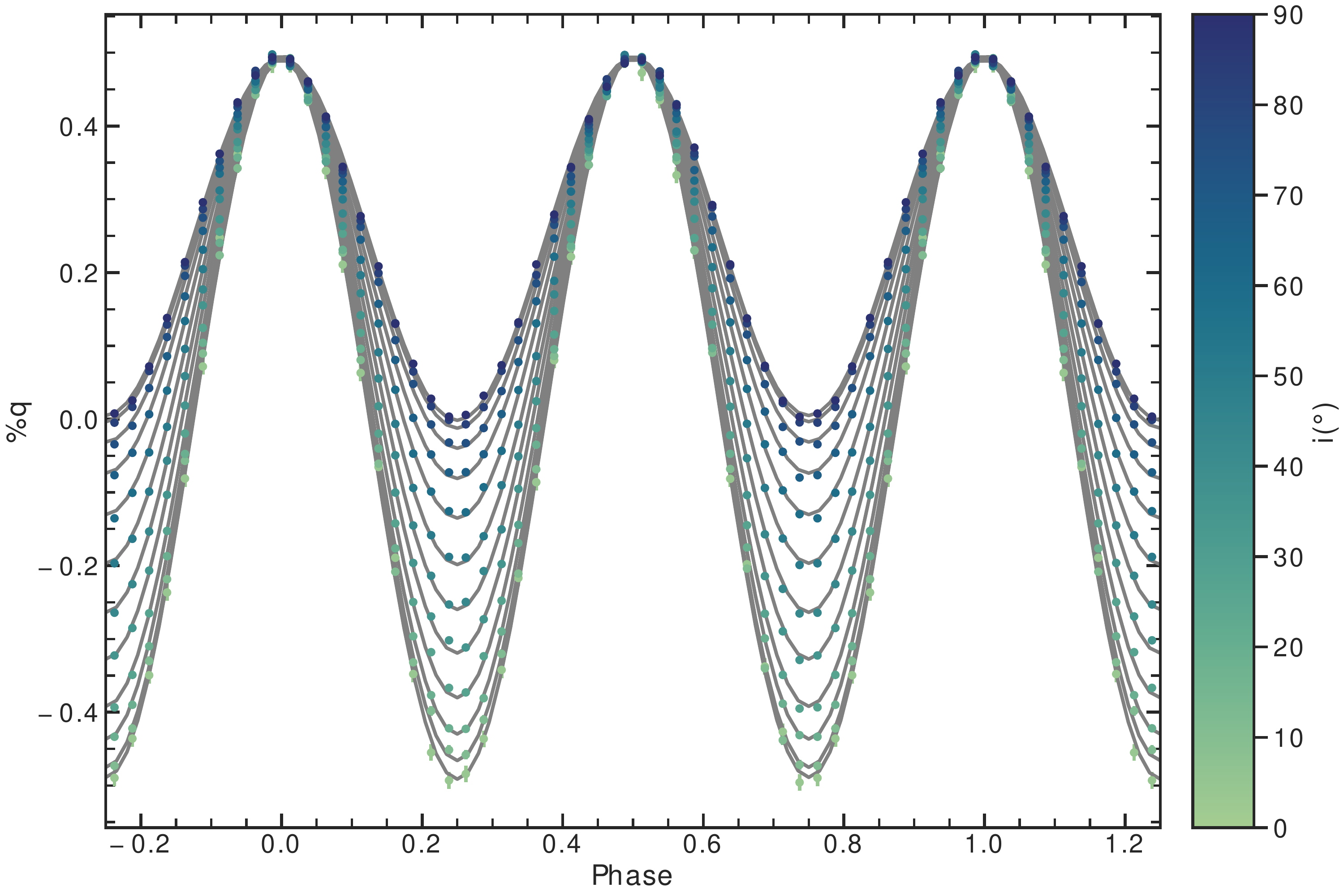}
\includegraphics[width=0.85\columnwidth]{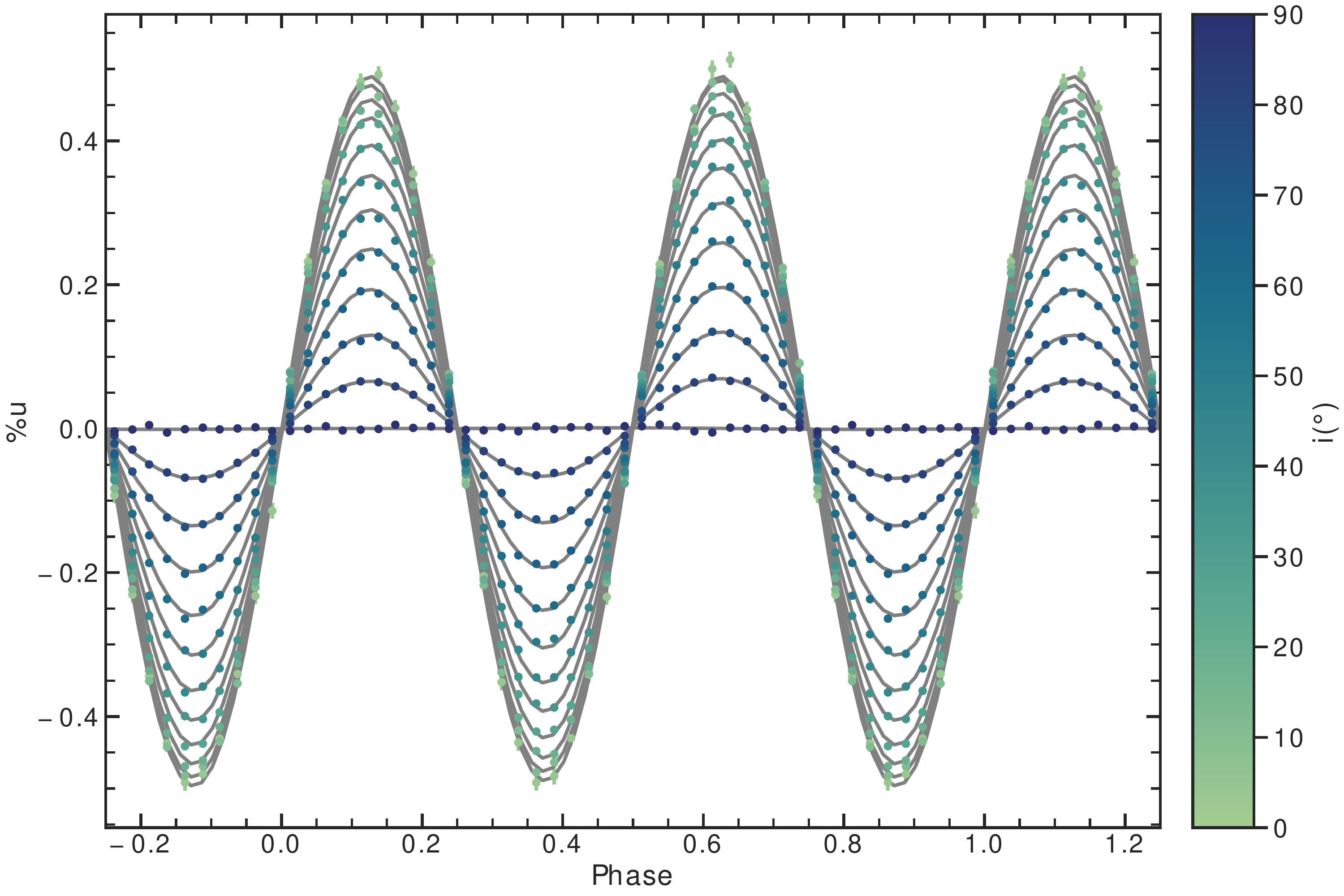}
\caption{Comparison of \textsc{slip} results for a sphere to the analytic model of \citet{brown_polarisation_1978}. The circles represent \textsc{slip} results for a range of inclination angles from $0-90\degr$ as displayed in the colour bar. The data have been fit by the model of \citet{brown_polarisation_1978} (grey line).}
\label{fig:bme_sphere_slip}
\end{figure}

\begin{figure}
\centering
\includegraphics[width=0.85\columnwidth]{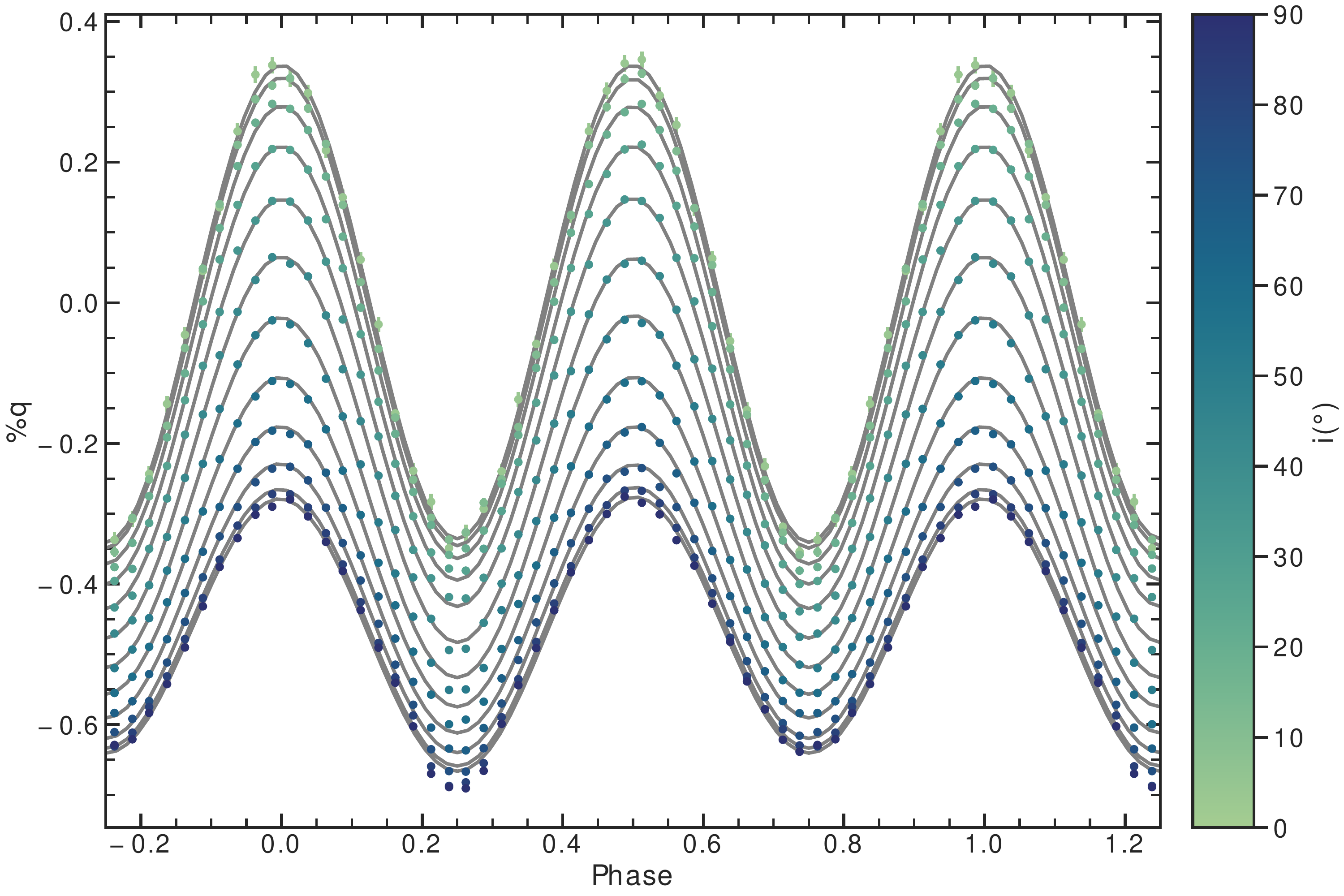}
\includegraphics[width=0.85\columnwidth]{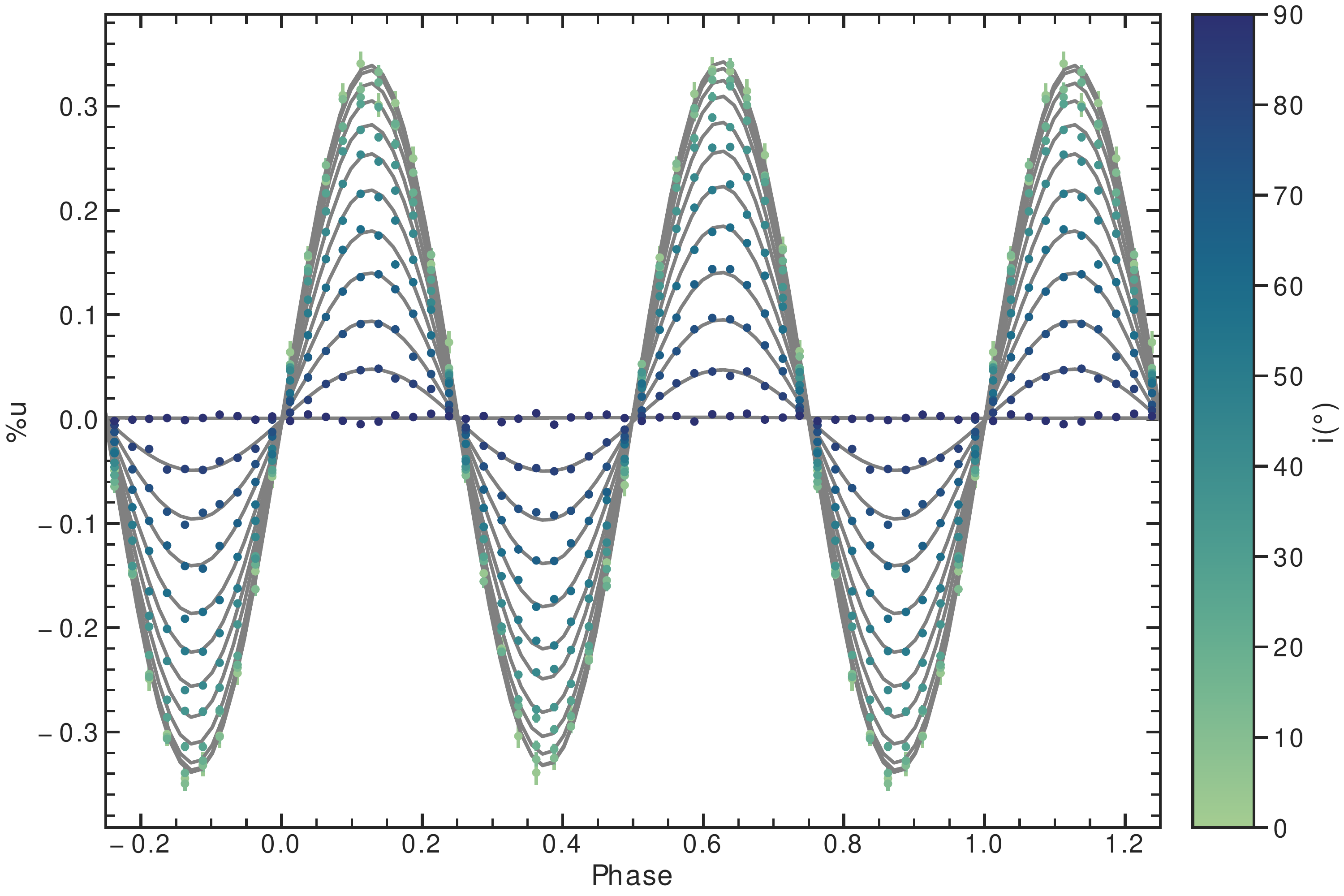}
\caption{Comparison of \textsc{slip} results for a prolate ellipsoid (extended in the direction perpendicular to the orbital plane) to the analytic model of \citet{brown_polarisation_1978}. The circles represent \textsc{slip} results for a range of inclination angles from $0-90\degr$ as displayed in the colour bar. The data have been fit by the model of \citet{brown_polarisation_1978} (grey line).}
\label{fig:bme_prolate_slip}
\end{figure}

The comparisons between the numerical and analytic models are shown in Figures \ref{fig:bme_disk_slip}, \ref{fig:bme_sphere_slip}, and \ref{fig:bme_prolate_slip} for the disk-like, spherical, and prolate density distributions, respectively. In all Figures~\ref{fig:bme_disk_slip}--\ref{fig:bme_prolate_slip}, the colour of the points corresponds to the inclination angle of the \textsc{slip} model. The uncertainty of the polarization increases as the inclination angle approaches zero (i.e. perpendicular to the orbital plane) because the amplitude of polarization is directly proportional to the orbital inclination angle \citep{brown_polarisation_1978}, and a lower amplitude increases the Monte Carlo noise in the simulation because fewer photons have a polarization signal. In Figure~\ref{fig:bme_disk_slip}, the significantly higher uncertainty of the \textsc{slip} model arises due to the extremely thin disk-like scattering region. This reduces the polarization from the model by an order of magnitude relative to the spherical and prolate density distributions, and thus the uncertainty is proportionally larger. 

\begin{figure*}
\gridline{
    \fig{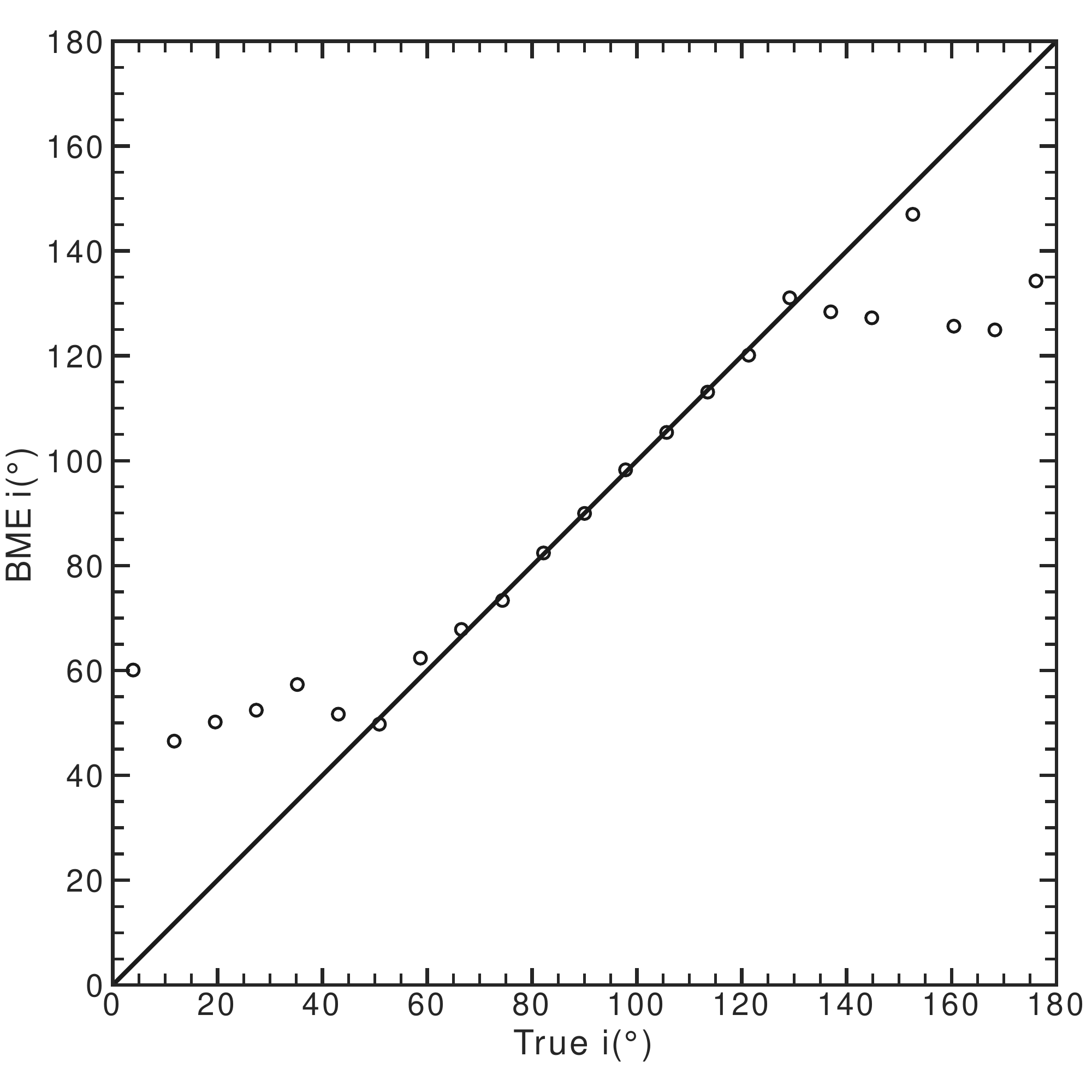}{0.3\columnwidth}{(a)}
    \fig{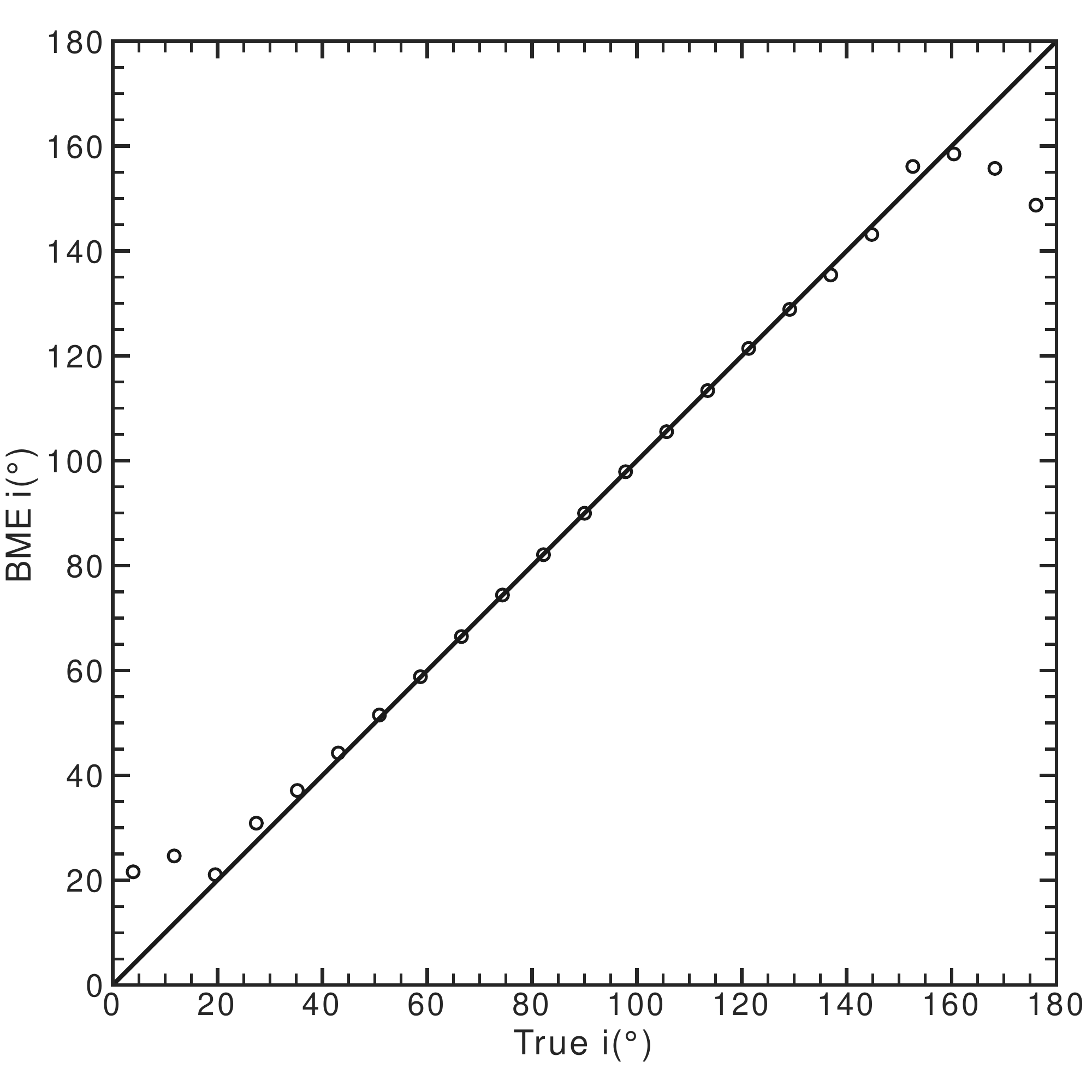}{0.3\columnwidth}{(b)}
    \fig{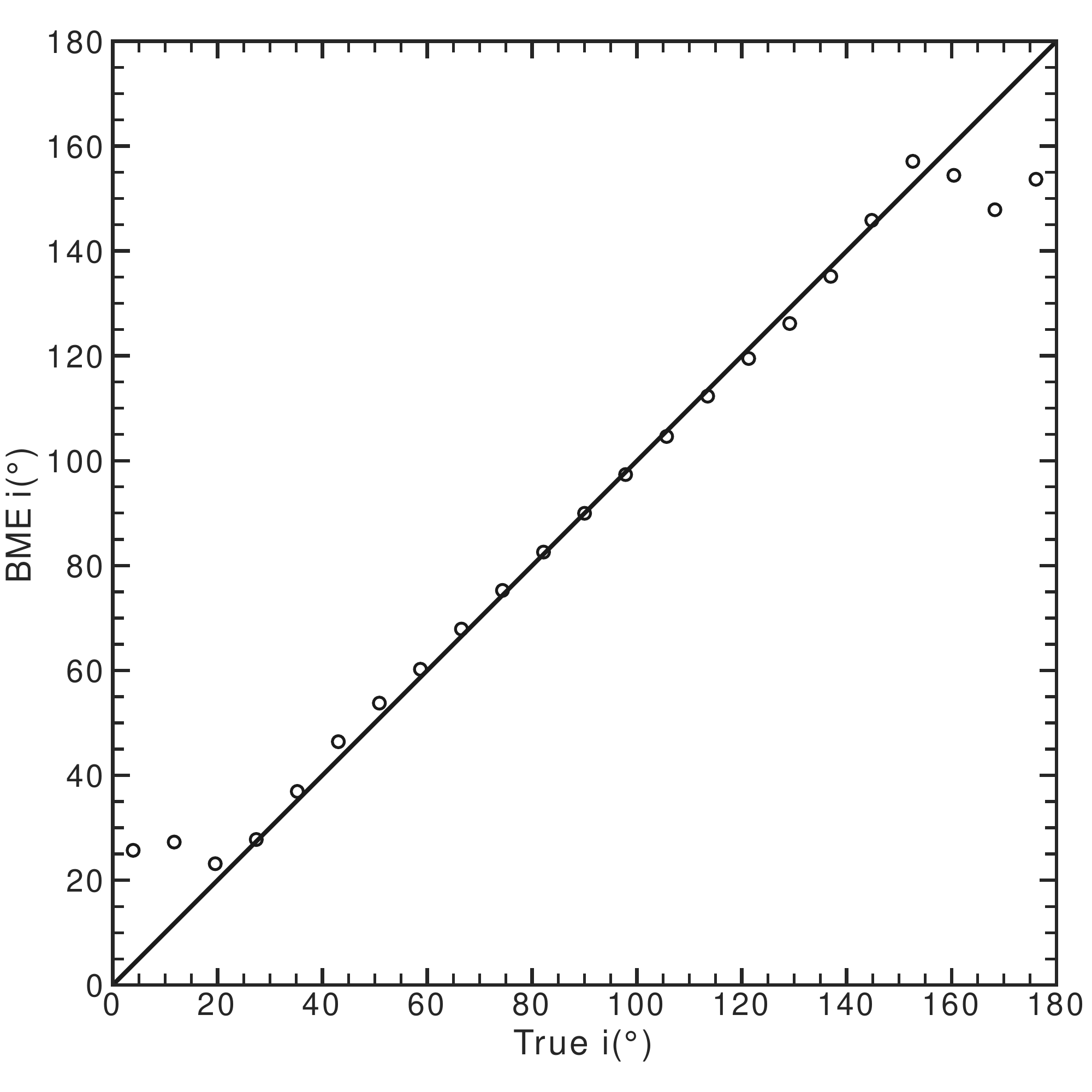}{0.3\columnwidth}{(c)}
}
\caption{Comparison of \textsc{slip} inclination for a thin disk (a), a sphere (b), and a prolate ellipsoid (c) distribution of scattering material to the analytic model of \citet{brown_polarisation_1978}. The points show the true \textsc{slip} model inclination versus the calculated \citeauthor{brown_polarisation_1978} model inclination. The black line shows unity.}
\label{fig:bme_inc}
\end{figure*}

The fit results for a selection of inclination angles are presented in Figure \ref{fig:bme_inc} for the thin disk, sphere, and prolate CSM models. As described by \citet{wolinski_confidence_1994}, the analytic model requires extremely high precision measurements relative to the polarization variation amplitude to recover accurate derived parameters as the inclination angle decreases from 90$\degr$ to 0$\degr$. Our results agree with those of \citet{wolinski_confidence_1994}: despite the high accuracy of the numerical models, the truly low inclination angle numerical models are fitted with a biased higher inclination angle by the analytic model. The disk model is most strongly affected because it has the lowest amplitude of polarization variation of the models. 

\section{Emulator details}\label{sec:emulator_app}

Here we present the details of our emulator, including its training set and architecture.

\subsection{Training set}\label{sec:training_set}

The data set used to train the neural network was produced using 4517 runs of the modified \textsc{slip} MCRT code using uniform random sampling of the parameter space described in Table~\ref{tab:sample_space}. 

To create the training set, we used the MSU high-performance computing cluster provided by the Institute of Cyber-Enabled Research. Each run took up to four hours across 8 CPUs with $2\times10^8$ photons per CPU. This photon count was chosen to produce uncertainties in $q$ and $u$ below $\sim0.05\%$ with a median uncertainty of $\sim0.03\%$. The resulting training set including all inclination angle bins has size \num{406530 x 40} because of the uniform sampling of $i$ in each model and because each model has 40 total $\phi$ (phase) bins.

\subsection{Emulator architecture}\label{sec:architecture}

The emulator is based on the proven DALEK emulator methods that have previously been applied to the radiative transfer code \textsc{tardis} \citep{kerzendorf_spectral_2014, kerzendorf_dalek_2021}. It was built using Tensorflow \citep{martin_abadi_tensorflow_2015} and scikit-learn \citep{pedregosa_scikit-learn_2011} tools to provide the neural network and data preprocessing respectively. 

The neural network architecture was investigated using a hyperparameter search of \num{2000} possible architectures on a cluster with 54 GPUs running Polyaxon, on-premise at the Volkswagen Group Machine Learning Research Lab. The resulting best architecture was chosen based on the total loss (here, using the mean squared error) and the number of steps required to train. The neural network consists of five hidden layers of width 30, with softplus activation (where $\mathrm{softplus}(x)= \log[\exp{x} + 1]$) and Glorot normal initialization \citep{glorot_understanding_2010}. No dropouts or layer normalization was deemed necessary by the hyperparameter search.

The data and model input parameters were scaled using the StandardScaler function of scikit-learn. The data were randomly split into 90\%/10\% train/validation sets. Optimal neural network architectures were those that had a minimal error on the validation set. The training was performed on an nVidia RTX 2080Ti graphics card with 12 GB of VRAM, and took approximately 19 hours to reach \num{20000} epochs. Of these \num{20000} steps, the training result with the lowest total validation set loss was chosen to represent the best result.

Two emulators were produced from the training set, one for Stokes $q$ and one for Stokes $u$. Validation of the emulator against random samples of the training set showed that the two emulators returned consistent model parameters independently of each other.




\section{Corner plots}\label{sec:corners}

The corner plots show the 6-dimensional posterior distributions of each parameter with 2-dimensional posterior distributions for each pair of parameters. These distributions are available at Zenodo LINK TBA.

\begin{figure*}
\centering
\includegraphics[width=\textwidth]{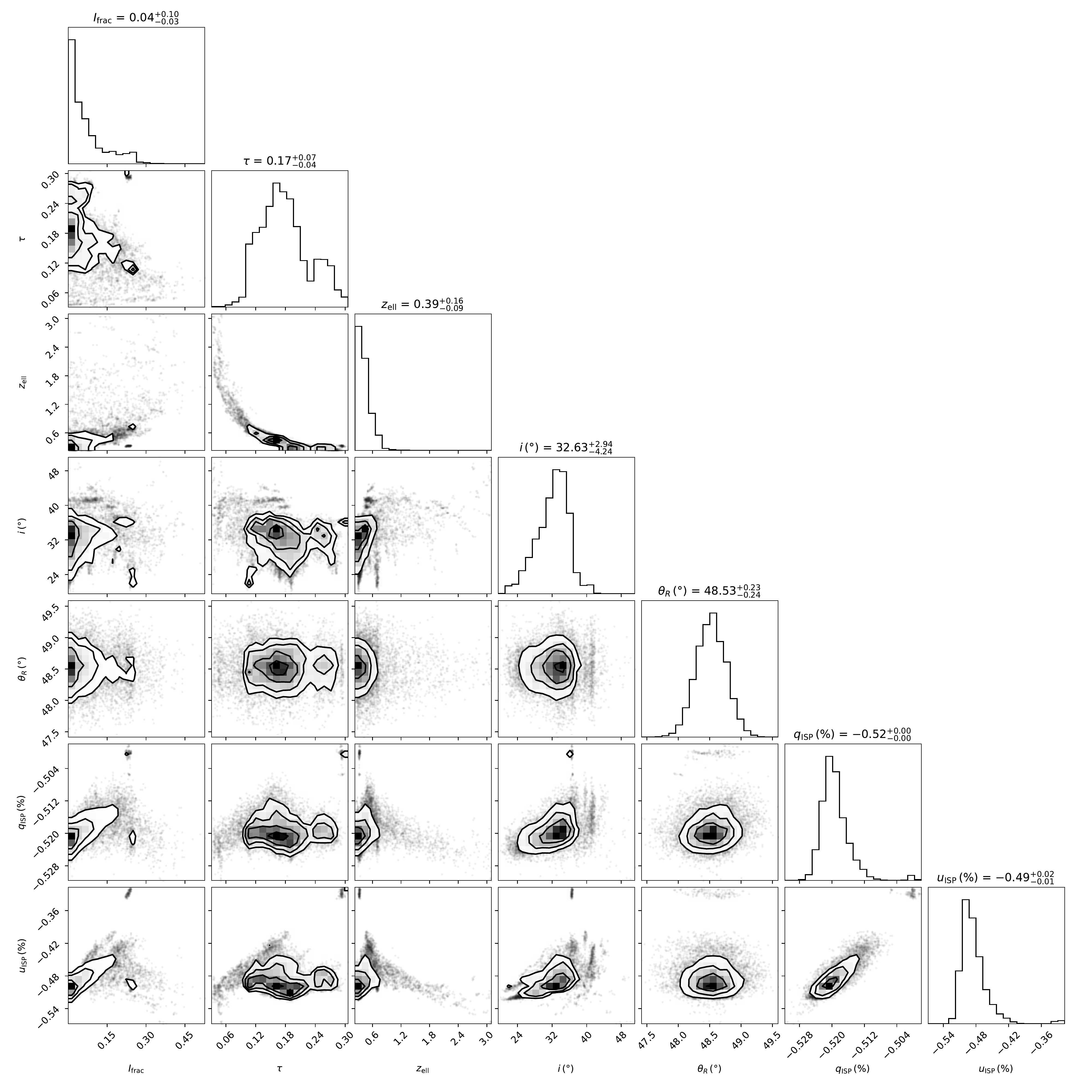}
\caption{Corner plot of WR 42 showing posterior distributions for each parameter and their relationships.}
\label{fig:corner_wr42_no_const}
\end{figure*}


\begin{figure*}
\centering
\includegraphics[width=\textwidth]{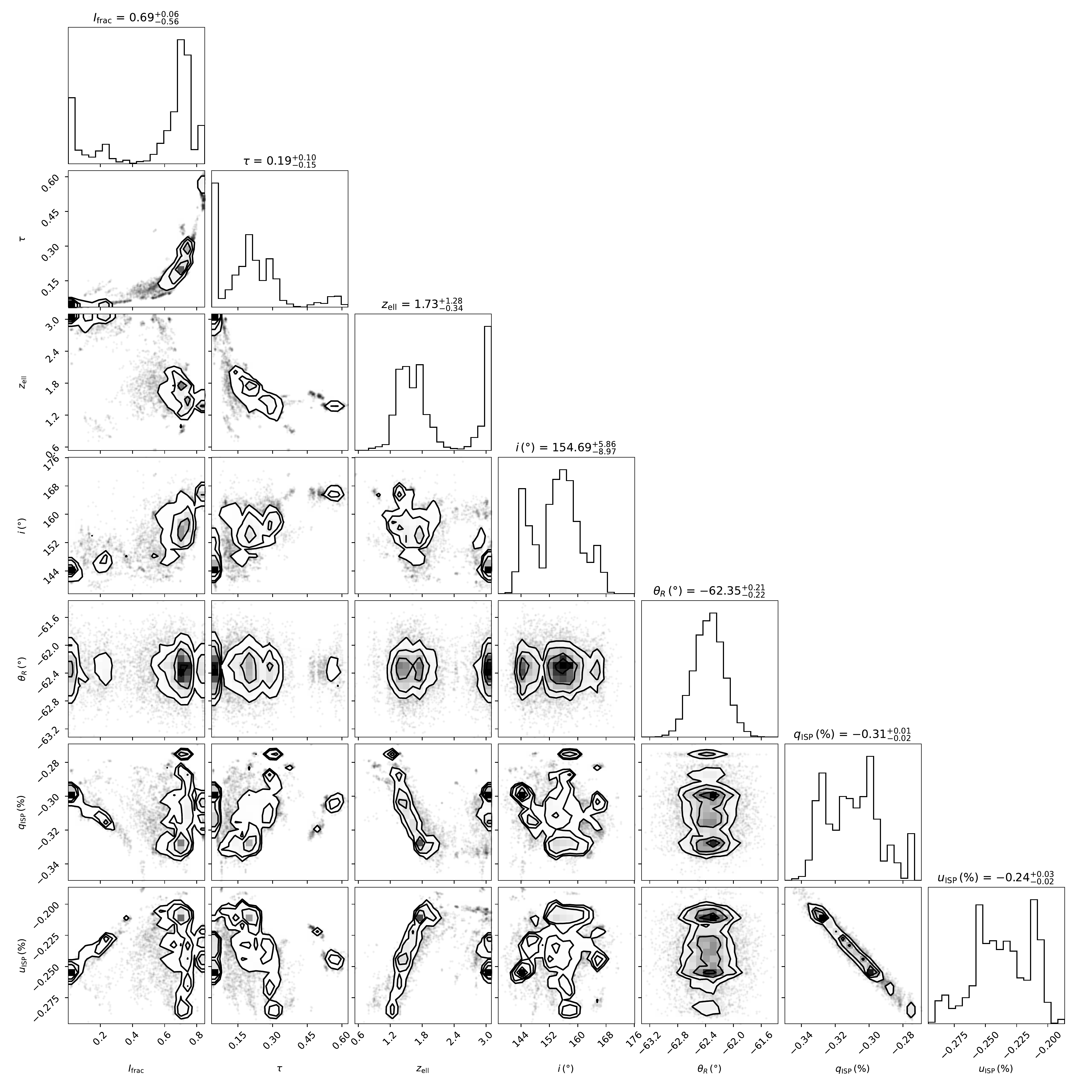}
\caption{Corner plot of WR 79 showing posterior distributions for each parameter and their relationships.}
\label{fig:corner_wr79}
\end{figure*}

\begin{figure*}
\centering
\includegraphics[width=\textwidth]{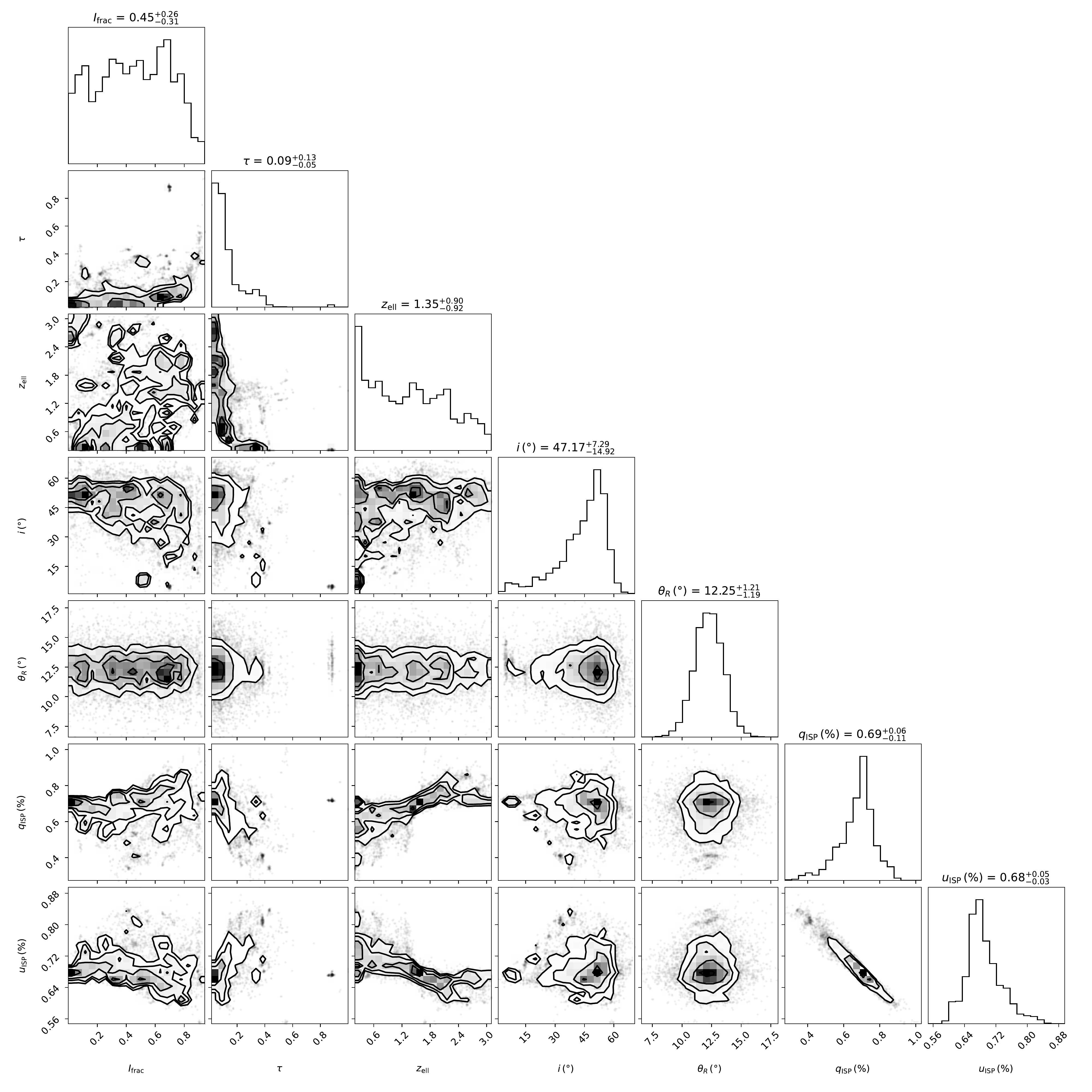}
\caption{Corner plot of WR 127 showing posterior distributions for each parameter and their relationships.}
\label{fig:corner_wr127}
\end{figure*}

\begin{figure*}
\centering
\includegraphics[width=\textwidth]{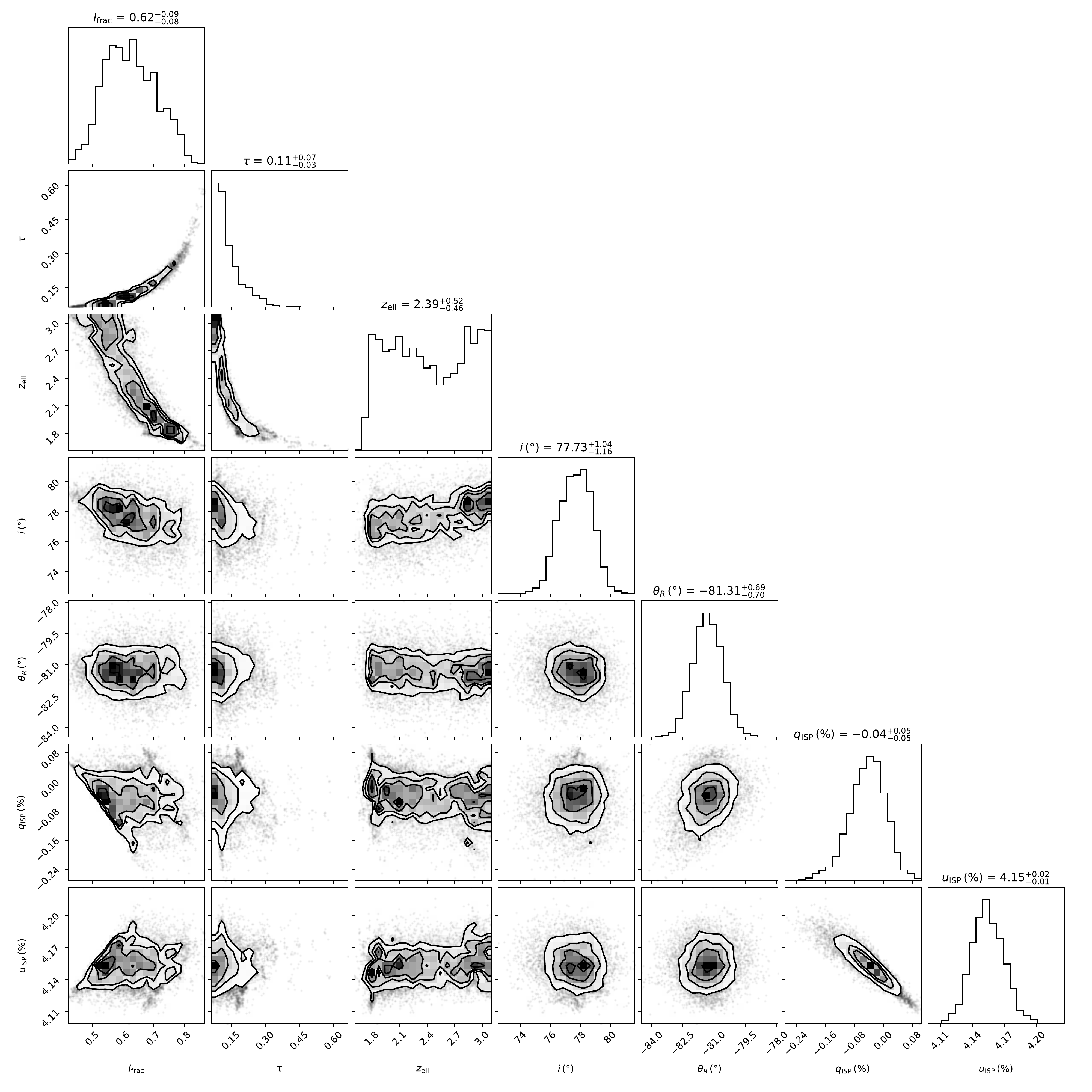}
\caption{Corner plot of WR 153 showing posterior distributions for each parameter and their relationships.}
\label{fig:corner_wr153}
\end{figure*}


\end{document}